\documentclass[
reprint,           
onecolumn,
superscriptaddress,
amsmath,           
amssymb,           
aps,               
prd,               
notitlepage,       
longbibliography,  
floatfix,          
nofootinbib,
]{revtex4-1}
\usepackage{amsmath}
\allowdisplaybreaks[4]
\usepackage{cancel}
\usepackage{extarrows}
\usepackage{tensor}     
\usepackage{float}
\usepackage[caption = false]{subfig} 
\usepackage[final]{graphicx}   
\usepackage[
colorlinks=true,        
citecolor=blue,         
linkcolor=blue,         
urlcolor=blue           
]{hyperref}             
\usepackage{bm}         
\usepackage{xcolor}     
\usepackage{lipsum}
\usepackage{color}      
\usepackage[utf8]{inputenc} 
\usepackage[section]{placeins} 
\usepackage{appendix}
\usepackage{units}
\newcommand{\nc}{\newcommand*}

\nc{\tr}{\rm{tr}}
\nc{\sig}{\sigma}
\nc{\om}{\omega}
\nc{\bt}{\beta}
\nc{\nb}{\nabla}
\nc{\eps}{\epsilon}
\nc{\epsS}{\epsilon_{\rm{S}}}
\nc{\epsV}{\epsilon_{\rm{V}}}
\nc{\epsT}{\epsilon_{\rm{T}}}
\nc{\osc}{\rm{osc}}
\nc{\DM}{\rm{TF}}
\nc{\FP}{\rm{FP}}
\nc{\eV}{\rm{eV}}
\nc{\figurewidth}{3.2in}
\nc{\xbar}{\bar{x}}
\nc{\rhoeq}{\rho_{\mathrm{eq}}}
\nc{\zeq}{z_{\mathrm{eq}}}
\nc{\tla}{\tilde{\lambda}}
\nc{\dt}{\delta}
\nc{\Dt}{\Delta}
\nc{\vj}{\hat{j}}
\nc{\vl}{\hat{l}}
\nc{\hx}{\hat{x}}
\nc{\hy}{\hat{y}}
\nc{\bj}{\bm{j}}
\nc{\mJ}{\mathcal{J}}
\nc{\mP}{\mathcal{P}}
\nc{\Msun}{M_\odot}
\nc{\app}{\approx}
\nc{\av}[1]{\langle #1 \rangle}
\nc{\eq}[1]{Eq.~\eqref{#1}}
\nc{\al}{\alpha}
\nc{\Xstar}{X_{\ast}}
\nc{\seq}{\sigma_{\mathrm{eq}}}
\nc{\fpbh}{f_{\mathrm{pbh}}}
\nc{\vth}{\hat{\theta}}
\nc{\vla}{\hat{\lambda}}
\nc{\vd}{\hat{d}}
\nc{\Mmin}{M_{\mathrm{min}}}
\nc{\rmd}{\mathrm{d}}
\nc{\mmin}{{m_{\mathrm{min}}}}
\nc{\mmax}{{m_{\mathrm{max}}}}
\nc{\mR}{\mathcal{R}}
\nc{\tmR}{\tilde{\mathcal{R}}}
\nc{\s}{\sigma}
\nc{\ogw}{\Omega_{\mathrm{GW}}}
\nc{\addref}{[\textcolor{red}{add ref}] }
\nc{\Om}{\Omega}
\nc{\gm}{\gamma}
\nc{\Gm}{\Gamma}
\nc{\gpcyr}{\mathrm{Gpc}^{-3}\,\mathrm{yr}^{-1}}
\nc{\Eq}[1]{Eq.~\eqref{#1}}
\nc{\Fig}[1]{Fig.~\ref{#1}}
\nc{\Table}[1]{Table~\ref{#1}}
\nc{\lvc}{LIGO/Virgo} 
\nc{\Sec}[1]{Sec.~\ref{#1}}
\nc{\eg}{\textit{e.g.~}}
\nc{\SNR}{\mathrm{SNR}}
\nc{\be}{\mathbf{\epsilon}}
\nc{\bn}{\mathbf{n}}
\nc{\bx}{\mathbf{x}}
\nc{\bk}{\mathbf{k}}
\nc{\bd}{\mathbf{d}}
\nc{\ba}{\mathbf{a}}
\nc{\bp}{\mathbf{p}}
\nc{\bnu}{\mathbf{\nu}}
\nc{\uni}{\mathrm{U}}
\nc{\logu}{\operatorname{\mathrm{log-U}}}
\nc{\RN}{\mathrm{RN}}
\nc{\BN}{\mathrm{BN}}
\nc{\GN}{\mathrm{GN}}
\nc{\mcN}{\mathcal{N}}
\nc{\GWB}{\mathrm{GW}}
\nc{\yr}{\mathrm{yr}}
\nc{\Am}{\mathcal{A}}
\nc{\Dm}{\mathcal{D}}
\nc{\Hm}{\mathcal{H}}
\nc{\sovast}{Soviet Ast.}
\nc{\hosc}{h_{\mathrm{osc}}}
\nc{\Posc}{\Psi_{\mathrm{osc}}}

\nc{\mrm}{\mathrm}
\nc{\BE}{B\scriptsize{AYES}\normalsize{E}\scriptsize{PHEM}\normalsize  }

\def\p{\partial}

\def\({\left(}
\def\){\right)}
\def\[{\left[}
\def\]{\right]}

\def\e{\begin{equation}}
\def\q{\end{equation}}
\def\m{\begin{eqnarray}}
\def\n{\end{eqnarray}}
\nc{\red}[1]{\textcolor{red}{#1}}
\begin{document}

\title{Pulsar timing residual induced by ultralight tensor dark matter}     

\author{Yu-Mei Wu}
\email{ymwu@ucas.ac.cn} 
\affiliation{School of Fundamental Physics and Mathematical Sciences, Hangzhou Institute for Advanced Study, UCAS, Hangzhou 310024, China}
\affiliation{School of Physical Sciences, University of Chinese Academy of Sciences, No. 19A Yuquan Road, Beijing 100049, China}
\affiliation{CAS Key Laboratory of Theoretical Physics, Institute of Theoretical Physics, Chinese Academy of Sciences, Beijing 100190, China}

\author{Zu-Cheng Chen}
\email{Corresponding author: zucheng.chen@bnu.edu.cn}
\affiliation{Department of Astronomy, Beijing Normal University, Beijing 100875, China}
\affiliation{Advanced Institute of Natural Sciences, Beijing Normal University, Zhuhai 519087, China}
\affiliation{Department of Physics and Synergistic Innovation Center for Quantum Effects and Applications, Hunan Normal University, Changsha, Hunan 410081, China}
\author{Qing-Guo Huang}
\email{Corresponding author: huangqg@itp.ac.cn}
\affiliation{School of Fundamental Physics and Mathematical Sciences, Hangzhou Institute for Advanced Study, UCAS, Hangzhou 310024, China}
\affiliation{School of Physical Sciences, 
    University of Chinese Academy of Sciences, 
    No. 19A Yuquan Road, Beijing 100049, China}
\affiliation{CAS Key Laboratory of Theoretical Physics, 
    Institute of Theoretical Physics, Chinese Academy of Sciences,
    Beijing 100190, China}



\begin{abstract}
Ultralight boson fields, with a mass around $10^{-23}\text{eV}$, are promising candidates for the elusive cosmological dark matter. These fields induce a periodic oscillation of the spacetime metric in the nanohertz frequency band, which is detectable by pulsar timing arrays. In this paper, we investigate the gravitational effect of ultralight tensor dark matter on the arrival time of radio pulses from pulsars. We find that the pulsar timing signal caused by tensor dark matter exhibits a different angular dependence than that by scalar and vector dark matter, making it possible to distinguish the ultralight dark matter signal with different spins.
Combining the gravitational effect and the coupling effect of ultralight tensor dark matter with standard model matter provides a complementary way to constrain the coupling parameter $\alpha$.
We estimate $\alpha \lesssim 10^{-6}\sim 10^{-5}$ in the mass range $m<5\times 10^{-23}\mathrm{eV}$ with current pulsar timing array.

\end{abstract}

	
\maketitle
	
	
\section{Introduction}
Numerous cosmological observations \citep{1980ApJ...238..471R,1982ApJ...261..439R,1976ApJ...204..668F,Massey:2010hh} have provided overwhelming support for the existence of dark matter, and precise analyses of cosmic microwave background (CMB) \citep{Planck:2018vyg} in particular have estimated the fraction of dark matter in the total energy of the Universe. However, the nature and properties of dark matter, such as its composition and interaction with the Standard Model (SM), remain mysterious. There are plenty dark matter candidates \citep{Bertone:2004pz,Feng:2010gw,Bertone:2018krk}, from tiny axion-like particles with mass around $\unit[10^{-22}]{\rm {eV}}$ \citep{Marsh:2015xka} to planet-sized massive compact halo objects (MACHOs) of $m\sim \Msun$ \citep{MACHO:2000qbb}. Among the bunch candidates, the weakly interactive massive particles (WIMPs) are expected to be highly desirable. However, despite considerable efforts in searching for them, no signal of WIMPs has yet been observed \citep{Schumann:2019eaa}.

In recent years, ultralight dark matter (ULDM) with a mass around $\unit[10^{-23}]{\rm{eV}}$ has attracted wide attention \citep{Hu:2000ke,Hui:2016ltb}. The ULDM is assumed to be condensed before inflation and oscillates coherently when its mass is larger than the Hubble scale \citep{Fox:2004kb}. On the one hand, ULDM moves at a non-relativistic speed, retaining the advantages of traditional cold dark matter to explain the large-scale structure of galaxies. On the other hand, its large de Broglie wavelength can suppress the structure on small scales, providing feasible solutions to the puzzles encountered by the traditional cold dark matter \citep{Hui:2016ltb,Niemeyer:2019aqm}, such as the core-cusp problem \cite{Gentile:2004tb,deBlok:2009sp} and missing-satellites problem \cite{Moore:1999nt,Klypin:1999uc}.

ULDM has been traditionally represented by scalar \citep{Hu:2000ke,Khmelnitsky:2013lxt,Hui:2016ltb} or pseduo-scalar fields (spin-0) such as axion-like particles \citep{Arvanitaki:2009fg, Hlozek:2014lca,Payez:2014xsa,Ivanov:2018byi}.
However, fields with higher spins, such as the massive vector fields (spin-1), 
can also act as alternative candidates for ULDM. Although ULDM candidates with different spins share many similarities in the cosmological phenomena, each possesses specific features. For instance, the massive vector particles can interact with SM particles if they are gauge bosons for $U(1)_{B}$ (``B" refers to baryon) or $U(1)_{B-L}$ (``L" refers to lepton) symmetry \citep{Pierce:2018xmy}, and their multiple polarization modes may cause anisotropies in certain experiments \citep{Nomura:2019cvc}.

Recently, massive tensor fields (spin-2) have been proposed as another viable candidate for dark matter \citep{Aoki:2016zgp,Babichev:2016hir,Babichev:2016bxi,Aoki:2017cnz}.
The tensor ULDM can be naturally predicted by a ghost-free bimetric theory \citep{Hassan:2011zd} in some mass range \citep{Marzola:2017lbt}.
This theory has two mass eigenstates at the linearized level: a massless spin-2 field $G_{\mu\nu}$ and a massive spin-2 field $M_{\mu\nu}$. At nonlinear levels, the expansion of the action \citep{Babichev:2016hir,Babichev:2016bxi,Marzola:2017lbt,Gialamas:2023aim} reveals that the self-interaction terms of the massless graviton $G_{\mu\nu}$ exactly match the Einstein-Hilbert structure. Meanwhile, the massive mode gravitates with the same strength as SM matter. But terms that contain one $M_{\mu\nu}$ and at least one $G_{\mu\nu}$, such as $MG$, $MG^2$, and $MG^3$, are missing in the expansion.
As a result, the terms with $G_{\mu\nu}$ can be partially resumed up to an effective background $g_{\mu\nu}$. In contrast, the massive fluctuation $M_{\mu\nu}$ can be treated as a propagating massive spin-2 field on the background $g_{\mu\nu}$.

The ULDM can leave imprints on cosmological and astrophysical observables. Its mass range and properties have been explored by many experiments, such as CMB \citep{Hlozek:2014lca,Poulin:2018dzj}, Lyman-$\alpha$ \citep{Irsic:2017yje,Irsic:2017yje}, non-observation of superradiance \citep{Unal:2020jiy}, secular variations in binary-pulsar orbital parameters \citep{Armaleo:2019gil,Blas:2019hxz}, etc.  
In particular, for dark matter fields with a mass around $\unit[10^{-23}]{\rm{eV}}$, pulsar timing arrays (PTAs) provide a powerful tool for exploring both its gravitational and non-gravitational properties \citep{Khmelnitsky:2013lxt,Porayko:2014rfa,Porayko:2018sfa,Kato:2019bqz,Nomura:2019cvc,PPTA:2021uzb,PPTA:2022eul,Armaleo:2020yml}. A PTA consists of a set of millisecond pulsars for which the arrival time of the emitted radio pulses are monitored with high precision for decades \citep{1978SvA....22...36S,Detweiler:1979wn,1990ApJ...361..300F}, recording the spacetime disturbances in the nanohertz range.
The coherent motion of ULDM carrying energy and momentum induces nontrivial oscillations of the metric \citep{Khmelnitsky:2013lxt,Nomura:2019cvc}, which can leave a mark on the timing residuals of PTA, known as the ``gravitational effect". Additionally, if ULDM interacts with SM particles, it can lead to pulsar spin fluctuations and reference clock shifts due to the induced variations in fundamental constants of the SM \citep{Kaplan:2022lmz} or can exert periodic forces on both the pulsar and Earth, resulting in relative displacements between them \citep{PPTA:2021uzb}. These effects are called the ``coupling effect" and can cause timing residuals. 

The specific form of the pulsar timing signal depends on the spin of ULDM. One unique aspect of tensor dark matter is that it naturally couples to SM fields without manually adding a fifth force to mimic the modification of gravity.
The coupling effect on PTA has been has been investigated in previous studies \citep{Armaleo:2020yml, Sun:2021yra, Unal:2022ooa}. Our present study instead focuses on providing the first comprehensive analysis of the gravitational effect of tensor dark matter on PTA. This analysis reveals distinct differences in angular dependence compared to scalar and vector dark matter due to the presence of more polarization modes. 
Furthermore, the gravitational effect also differs from the coupling effect regarding angular dependence. The gravitational effect influences the arrival time of pulses through metric fluctuations induced by the energy-momentum tensor, which is quadratic in the tensor fields. In contrast, the coupling effect depends on the tensor fields linearly.
The paper is organized as follows. In Section \ref{sec2}, we show the time-dependent spacetime perturbations induced by the oscillating tensor ULDM. In Section \ref{sec3}, we derive the pulsar timing residuals induced by the tensor ULDM from the gravitational effect. We also discuss the constraints that can be inferred from current data and compare the gravitational and coupling effects in \ref{sec4}. In Section \ref{sec5}, we summarize the results.

\section{Metric perturbations induced by tensor ULDM}\label{sec2}
The action for a massive spin-2 filed $M_{\mu\nu}$ propagating on a background $g_{\mu\nu}$ can be formally written as \citep{Marzola:2017lbt}
\e
S=\frac{M_{\rm{P}}^2}{2}\int d^4 x \sqrt{-g}R(g)+\int d^4 x \sqrt{-g}  {\cal{L}}_{\FP}^{(2)}(M)+{\cal{O}}(M_{\mu\nu}^3),
\label{action}
\q
where $M_{\rm{P}}$ is the reduced Planck mass, and ${\cal{L}}_{\FP}^{(2)}(M)$ is the quadratic Fierz-Pauli Lagrangian,
\e
{\cal{L}}_{\FP}^{(2)}(M)=\frac{1}{8} \[-\nb_{\rho}M_{\mu\nu}\nb^{\rho}M^{\mu\nu}+2\nb_{\rho}M_{\mu\nu}\nb^{\nu}M^{\mu\rho}-2\nb_{\nu}M^{\mu\nu}\nb_{\mu}M+\nb^{\mu}M\nb_{\mu}M+m^2(M^2-M_{\mu\nu}M^{\mu\nu})\]
\q
with  the trace $M=M_{\,\,\,\mu}^{\mu}$. 
The cosmic expansion is negligible on galactic scales, so we treat the background to be flat. The equation of motion of the tensor fields can be derived from the action \Eq{action} by taking the variation with respect to $M_{\mu\nu}$,
\e\label{EOM1}
\square M^{\mu\nu}-(\p^{\mu}\p_{\rho}M^{\nu\rho}+\p^{\nu}\p_{\rho}M^{\mu\rho})+\eta^{\mu\nu}\p_{\rho\sig}M^{\rho\sig}+\p^{\mu\nu}M-\eta^{\mu\nu}\square M=m^2(M^{\mu\nu}-\eta^{\mu\nu}M).
\q
Applying $\p_{\mu}$ to both sides of \Eq{EOM1}, the left-hand side gives zero identically, so we obtain
\e\label{trk1}
m^2\p_{\mu}(M^{\mu\nu}-\eta^{\mu\nu}M)=0.
\q
Further, taking the trace of both sides of \Eq{EOM1}, we get
\e\label{trk2}
2\p_{\mu}\p_{\nu}(M^{\mu\nu}-\eta^{\mu\nu}M)=-3m^2 M.
\q
With \Eq{trk1} and \Eq{trk2}, we can rewrite \Eq{EOM1} into a set of equations for a massive tensor field in the standard form
\m
M&=&0,\label{trless}\\
\p_{\mu}M^{\mu\nu}&=&0,\label{HC}\\
(\square-m^2)M^{\mu\nu}&=&0. \label{EOM2}
\n

Now we consider the scenario where the spin-2 fields with an ultralight mass act as the dark matter following the discussion in Refs.~\citep{Khmelnitsky:2013lxt, Nomura:2019cvc, Armaleo:2020yml}. For ULDM with the typical mass $m\sim 10^{-23}\rm{eV}$ and the typical velocity in the galaxy $v\sim 10^{-3}$, the occupation number can be estimated to be as high as $10^{95}$, allowing us to treat the ultralight field as a classical wave. The wave is characterized by a typical wave number $k=m v$ and frequency $\omega\approx m+m v^2/2\approx m$ due to the non-relativistic velocity. Additionally, the de Broglie wavelength of the ULDM can be calculated as 
\e
\lambda_{\rm{dB}}=\frac{2\pi}{k}\approx 4{\rm{kpc}}\(\frac{10^{-23}{\eV}}{m}\)\(\frac{10^{-3}}{v}\),
\q
indicating the spin-2 field oscillates coherently with a monochromatic frequency determined by its mass on the scale of $\lambda_{\rm{dB}}$.

We return to Eqs.~(\Ref{trless}-\Ref{EOM2}). Note that the constraint in \Eq{HC} indicates that $M^{00}$ is suppressed by the order of $k/m=v\sim 10^{-3}$ compared to $M^{0j}$, while $M^{0j}$ is suppressed by the same order compared to $M^{ij}$.
Thus, we can neglect all the time components $M^{0\mu}$. Since the remains six spatial components $M_{ij}$ are subject to the traceless constraint in \Eq{trless}, the solution to \Eq{EOM2} is given by
\e
M_{ij}(t, \bx)=A(\bx) \cos(mt +\al(\bx))\eps_{ij}.
\q
We have neglected the spatial derivatives when solving the equation but retain the spatial dependence of the amplitude $A(\bx)$ and the phase $\alpha(\bx)$. The symmetric, traceless matrix $\eps_{ij}$ with a unit norm encoding the five free degrees of freedom of the massive spin-2 fields is \citep{Armaleo:2019gil},
\m
\begin{gathered}
\eps_{ij}=\frac{1}{\sqrt{2}}\begin{pmatrix}
\epsT\cos\chi-\epsS/\sqrt{3} & \epsT \sin \chi & \epsV \cos \xi \\
\epsT \sin \chi & -\epsT\cos\chi-\epsS/\sqrt{3} & \epsV \sin \xi\\
\epsV \cos \xi & \epsV \sin \xi & 2\epsS/\sqrt{3}
\end{pmatrix}
\end{gathered}.
\n
The amplitude parameters $\epsS$, $\epsV$, and $\epsT$, satisfying $\epsS^2+\epsV^2+\epsT^2=1$, represent the ratios of the scalar, vector, and tensor sector of the massive spin-2 fields, respectively. Meanwhile, the angle parameters $\xi$ and $\chi$ determine the azimuthal direction of the helicity-$\pm 1$ and helicity-$\pm 2$ modes, respectively. 
%

Now let's focus on the metric perturbation induced by the oscillating fields. Using the definition of energy-momentum from the matter fields, 
\e
T_{\mu\nu}=\frac{-2}{\sqrt{-g}}\frac{\dt S}{\dt g^{\mu\nu}},
\q
the energy-momentum of the massive spin-2 fields in a flat background is given by, 
\e
T_{\mu\nu}=-\frac{1}{8}\[\eta_{\mu\nu}\(\p_{\lambda}M^{\gm\dt}\p^{\lambda}M_{\gm\dt}+ m ^2 M^{\gm\dt}M_{\gm\dt}\)-2\p_{\mu}M_{\gm\dt}\p_{\nu}M^{\gm\dt}-4\p_{\lambda}M_{\mu\dt}\p^{\lambda}M^{\dt}_{\,\,\nu}-4m^2M_{\mu\dt}M^{\dt}_{\,\,\nu}\].
\q
with the corresponding time-dependent energy density
\e
T_{00}=\frac{1}{8}m^2 A(\bx)^2=\rho_{\rm{TF}}(\bx),
\label{T00}
\q
and the oscillating spatial components
\small
\m\label{Tij}
\begin{gathered}
T_{ij}=\frac{1}{24}\!m^2 \!A(\bx)^2 \!\cos(2m t\!+\!2\al(\bx))\qquad\qquad\qquad\qquad\qquad\qquad\qquad\qquad\qquad\qquad\qquad\qquad\qquad\qquad\qquad\qquad\qquad\qquad\\
\times\begin{pmatrix}\!-(\epsS^2\!+4\sqrt{3}\epsS \epsT c_{\chi}\!-3(\epsT^2\!+\epsV^2 c_{2\xi})) & \!-4\sqrt{3}\epsS \epsT s_{\chi}\!+3\epsV^2 s_{2\xi} & 2\epsV (\sqrt{3} \epsS  c_{\xi}\!+3\epsT  c_{\chi-\xi})   \\
\!-4\sqrt{3}\epsS \epsT s_{\chi}\!+3\epsV^2 s_{2\xi} & -(\epsS^2\!-4\sqrt{3}\epsS \epsT c_{\chi}\!-3\epsT^2\!+3\epsV^2 c_{2\xi})& 2 \epsV(\sqrt{3} \epsS  s_{\xi}\!+3\epsT s_{\chi-\xi})\\
2\epsV (\sqrt{3} \epsS  c_{\xi}+3\epsT c_{\chi-\xi}) & 2\epsV(\sqrt{3} \epsS  s_{\xi}+3\epsT s_{\chi-\xi}) & 5\epsS^2-3\epsT^2+3\epsV^2
\end{pmatrix}.
 \\
\quad \!\!\!\!\!\!\!\!\!
\end{gathered}
\n
\normalsize
Here we have used the notation $c_{x}=\cos x$ and $s_{x}=\sin x$ following Ref.~\citep{Armaleo:2019gil}. Note that although the anisotropic pressure $T_{ij}$ oscillate with the frequency, 
\e
\om=2 m,
\q
it averages to zero on the cosmological time scales, much larger than the oscillation period. Thus, the coherently oscillating spin-2 massive fields can be considered the pressureless non-relativistic matter. However, as shown below, the oscillating spatial components induce oscillation in spacetime metric, leading to observable effects in PTAs.

The oscillating massive spin-2 fields generate disturbance in the galactic-scale flat metric, which can be expressed as
\e\label{met1}
ds^2=-\[1+2\Phi(t,\bx)\]dt^2+\[1+2\Psi(t,\bx)\]\dt_{ij}dx^{i}dx^{j}+h_{ij}(t,\bx)dx^{i}dx^{j},
\q
where the perturbations are described by $\Phi$, $\Psi$ and $h_{ij}$. The spatial perturbation has been particularly decomposed into a trace part $\Psi$ and a traceless part $h_{ij}$ as the spatial components of the energy-momentum are anisotropic. Accordingly, we also decompose the symmetric  $T_{ij}$ into 
\e
T_{ij}=\frac{1}{3}\dt_{ij}T^{k}_{\,\,k}+\(T_{ij}-\frac{1}{3}\dt_{ij}T^{k}_{\,\,k}\),
\q
where the first term corresponds the trace part and the second term corresponds to the traceless part. From \Eq{Tij}, we obtain
\e
T^{k}_{\,\,k}=\frac{1}{8}m^2A(\bx)^2\cos(2mt +2\al(\bx)),
\q
and 
\small
\m
\begin{gathered}
\qquad T_{ij}-\frac{1}{3}\dt_{ij}T^{k}_{\,\,k}\qquad\qquad\qquad\qquad\qquad\qquad\qquad\qquad\qquad\qquad\qquad\qquad\qquad\qquad\qquad\qquad\qquad\qquad\qquad\qquad\qquad\qquad\\
=\frac{1}{24}m^2\!A(\bx)^2\cos(2mt\! +\!2\al(\bx))\qquad\qquad\qquad\qquad\qquad\qquad\qquad\qquad\qquad\qquad\qquad\qquad\qquad\qquad\qquad\qquad\qquad\qquad\qquad\\
\times\begin{pmatrix}
-(2\epsS^2\!-2\epsT^2\!+\epsV^2\!+4\sqrt{3}\epsS\epsT c_{\chi}\!-3\epsV^2 c_{2\xi}) &
-4\sqrt{3}\epsS\epsT s_{\chi}\!+3\epsV^2 s_{2\xi} &  2\epsV (\sqrt{3} \epsS  c_{\xi}\!+3\epsT  c_{\chi\!-\!\xi})   \\
\!-4\sqrt{3}\epsS \epsT s_{\chi}\!+3\epsV^2 s_{2\xi} & -(2\epsS^2\!-2\epsT^2\!+\epsV^2\!-4\sqrt{3}\epsS\epsT c_{\chi}\!+3\epsV^2 c_{2\xi})&2 \epsV(\sqrt{3} \epsS  s_{\xi}\!+3\epsT s_{\chi\!-\!\xi})\\
2\epsV (\sqrt{3} \epsS  c_{\xi}\!+3\epsT c_{\chi\!-\xi}) & 2\epsV(\sqrt{3} \epsS  s_{\xi}\!+3\epsT s_{\chi\!-\xi}) & 2(2\epsS^2-2\epsT^2\!+\epsV^2)\\
\end{pmatrix}.
\\
\end{gathered}
\n
\normalsize

Now we can solve the metric perturbations in \Eq{met1}. We will start with the trace part. In the Newtonian gauge, the metric is given by
\e\label{met}
ds^2=-(1+2\Phi(t,\bx))dt^2+(1+2\Psi(t,\bx))\dt_{ij}dx^{i}dx^{j}.
\q
We can split the gravitational potentials $\Phi$ and $\Psi$ into a dominant time-independent part and a time-dependent part oscillating with the frequency $\om$ as follows,
\m\label{osc1}
\Phi(t,\bx)&\approx& \Phi_0(\bx)+\Phi_{\osc} \cos(\om t +2\al(\bx)),\notag\\
\Psi(t,\bx)&\approx& \Psi_0(\bx)+\Psi_{\osc} \cos(\om t +2\al(\bx)),
\n
where $\Phi_{\text{osc}}$ and $\Psi_{\text{osc}}$ are the oscillation amplitudes. 
We can determine the time-independent part $\Psi_0(\mathbf{x})$ by solving the $00$ component of the linearized Einstein equations,
\e
\p_i \p^{i} \Psi=-4\pi G T_{00}=-4\pi G \rho_{\rm{TF}}(\bx).
\q
The oscillating part of $\Psi$ can be found by taking the trace of the spatial components of the linearized Einstein equations,
\e
-3\ddot{\Psi}+\p_i \p^{i}(\Phi+\Psi)=4\pi G T^{k}_{\,\,k},
\q
of which the time-independent part implies that $\Phi_{0}=-\Psi_{0}$. Therefore, the oscillation amplitude $\Psi_{\osc}$ is 
\e
\Psi_{\osc}=\frac{1}{24}\pi G A^2(\bx)=\frac{\pi G \rho_{\rm{TF}}(\bx)}{3m^2}.
\q

Next, we turn to the traceless part of the metric perturbation, $h_{ij}$. The metric, in this case, is given by
\e
ds^2 = -dt^2+\[\dt_{ij}+h_{ij}(t,\bx)\] dx^i dx^j,
\q
with $h^i_{\,\,i}=0$. After neglecting the suppressed spatial gradients $\p_{k}^2 h_{ij}$, the linearized Einstein equation yields
\e
\ddot{h}_{ij}=16\pi G \(T_{ij}-\frac{1}{3}\dt_{ij}T^{k}_{\,\,k}\),
\q
 The traceless part of the metric perturbation can then be obtained as
\small
\m\label{osc2}
\begin{gathered}
h_{ij}(t,\bx)=-\frac{1}{6}\pi G A^2(\bx)\cos(2 mt +2\al(\bx))\qquad\qquad\qquad\qquad\qquad\qquad\qquad\qquad\qquad\qquad\qquad\qquad\qquad\qquad\qquad\qquad\qquad\quad \\
\qquad\quad\times\begin{pmatrix}
-(2\epsS^2\!-2\epsT^2\!+\epsV^2\!+4\sqrt{3}\epsS\epsT c_{\chi}\!-3\epsV^2 c_{2\xi}) &
-4\sqrt{3}\epsS\epsT s_{\chi}\!+3\epsV^2 s_{2\xi} &  2\epsV (\sqrt{3} \epsS  c_{\xi}\!+3\epsT  c_{\chi\!-\!\xi})   \\
\!-4\sqrt{3}\epsS \epsT s_{\chi}\!+3\epsV^2 s_{2\xi} & -(2\epsS^2\!-2\epsT^2\!+\epsV^2\!-4\sqrt{3}\epsS\epsT c_{\chi}\!+3\epsV^2 c_{2\xi})&2 \epsV(\sqrt{3} \epsS  s_{\xi}\!+3\epsT s_{\chi\!-\!\xi})\\
2\epsV (\sqrt{3} \epsS  c_{\xi}\!+3\epsT c_{\chi\!-\xi}) & 2\epsV(\sqrt{3} \epsS  s_{\xi}\!+3\epsT s_{\chi\!-\xi}) & 2(2\epsS^2-2\epsT^2\!+\epsV^2)
\end{pmatrix}.
\end{gathered}
\n
\normalsize
Similar to the amplitude of the oscillation in the trace part $\Psi_{\rm{osc}}$, we define the amplitude of the oscillation in the traceless part by
\m
h_{\osc}(\bx)\equiv \frac{1}{6}\pi G A^2(\bx)=\frac{4\pi G \rho_{\rm{TF}}(\bx)}{3m^2}.
\label{eq_h_osc}
\n
\normalsize

\section{Timing residual induced by tensor ULDM with gravitational effect}\label{sec3}
The time-dependent metric, \Eq{osc1} and \Eq{osc2}, induced by the oscillating fields will cause a redshift and time delay for any propagating signal. One such signal is the radio pulse emitted by millisecond pulsars, which can be measured with extremely high timing precision and provide a natural tool for probing oscillating ULDM. In the following, we will derive the timing residual induced by tensor ULDM considering the gravitational effect.

Consider a pulse propagating from a pulsar to the Earth,
the redshift is defined as the relative frequency change of the pulse,
\e
z(t)=\frac{\om_0-\om_{\rm{obs}}(t)}{\om_0},
\q
where $\om_0$ is the angular frequency of the pulse when emitting and $\om_{\rm{obs}}(t)$ is the observed angular frequency at the Earth at time $t$. Integrating the redshift gives the arrival time of the pulse
\e
R(t)=\int_0^t dt\, z(t').
\q

Solving the photon geodesic equation under the perturbed metric provides the frequency shift. 
For the trace part and traceless part, the corresponding redshifts are \citep{Nomura:2019cvc},
\begin{align}
z_{\Psi}(t)&=\Psi(t,\bx_{e})-\Psi(t-|\bx_{p}|,\bx_{p}), \label{z_psi} \\ 
z_h(t)&=\frac{1}{2}\hat{p}^i \hat{p}^j [h_{ij}(t,\bx_{e})-h_{ij}(t-|\bx_{p}|,\bx_{p})],
\label{z_h}
\end{align}
where $\hat{\bf{p}}=(\sin \theta \cos \phi, \sin \theta \sin \phi, \cos \theta)$ denotes a unit vector pointing from the Earth to the pulsar,
and $\bx_{e}$ and $\bx_{p}$ represent the location of the Earth and the pulsar, respectively. Since most monitored pulsars are located relatively close to the Earth (within 2 kpc), we assume that the tensor ULDM has the same oscillation amplitude $A$ at both the Earth and the pulsar. The total redshift is given by
\m
z_t(t)&=&z_{\Psi}(t)+z_h(t)\notag \\
&=&\frac{1}{24}\pi G A^2\[\epsS^2(-1-6 \cos2\theta)+\epsT^2(3+6\cos2\theta)-3\epsV^2\cos2\theta+8\sqrt{3}\epsS\epsT\cos(2\phi-\chi)\sin^2\theta\right.\notag\\
&&\left.\qquad\quad\,\,-2\epsV\(2\sqrt{3}\epsS\cos(\phi-\xi)\sin2\theta+6\epsT\cos(\phi-\chi+\xi)\sin2\theta+3\epsV \cos2(\phi-\xi)\sin^2\theta\)\]\notag\\
&&\times \[\cos(2mt+2\al(\bx_{e}))-\cos\(2m(t-|\bx_{p}|)+2\al(\bx_{p})\)\]\notag \\
&=&-\frac{2\pi G \rho_{\rm{TF}}}{3m^2}\[\epsS^2(-1-6 \cos2\theta)+\epsT^2(3+6\cos2\theta)-3\epsV^2\cos2\theta+8\sqrt{3}\epsS\epsT\cos(2\phi-\chi)\sin^2\theta\right.\notag\\
&&\left.\qquad\quad\,\,-2\epsV\(2\sqrt{3}\epsS\cos(\phi-\xi)\sin2\theta+6\epsT\cos(\phi-\chi+\xi)\sin2\theta+3\epsV \cos2(\phi-\xi)\sin^2\theta\)\]\notag\\
&& \times \sin (\alpha_{e}-\alpha_{p}) \sin \left(2 m t+\alpha_{e}+\alpha_{p}\right),
\label{z_t}
\n
where we defined the phases in Earth term and pulsar term as $\alpha_{e}=\alpha(\bx_{e})$ and $\alpha_{p}=\alpha\left(\bx_{p}\right)-m\left|\mathbf{x}_{p}\right|$, respectively.
Correspondingly, the timing residuals read,
 \m
 R_t&=&
\frac{\pi G \rho_{\rm{TF}}}{3m^3}
\[\epsS^2(-1-6 \cos2\theta)+\epsT^2(3+6\cos2\theta)-3\epsV^2\cos2\theta+8\sqrt{3}\epsS\epsT\cos(2\phi-\chi)\sin^2\theta\right.\notag\\
&&\left.\qquad\,\,-2\epsV\(2\sqrt{3}\epsS\cos(\phi-\xi)\sin2\theta+6\epsT\cos(\phi-\chi+\xi)\sin2\theta+3\epsV \cos2(\phi-\xi)\sin^2\theta\)\]\notag\\
&& \times \sin (\alpha_{e}-\alpha_{p}) \cos \left(2 m t+\alpha_{e}+\alpha_{p}\right).
\label{R_t}
 \n

\begin{figure}[htbp!]
	\centering
	\includegraphics[width=1.0\textwidth]{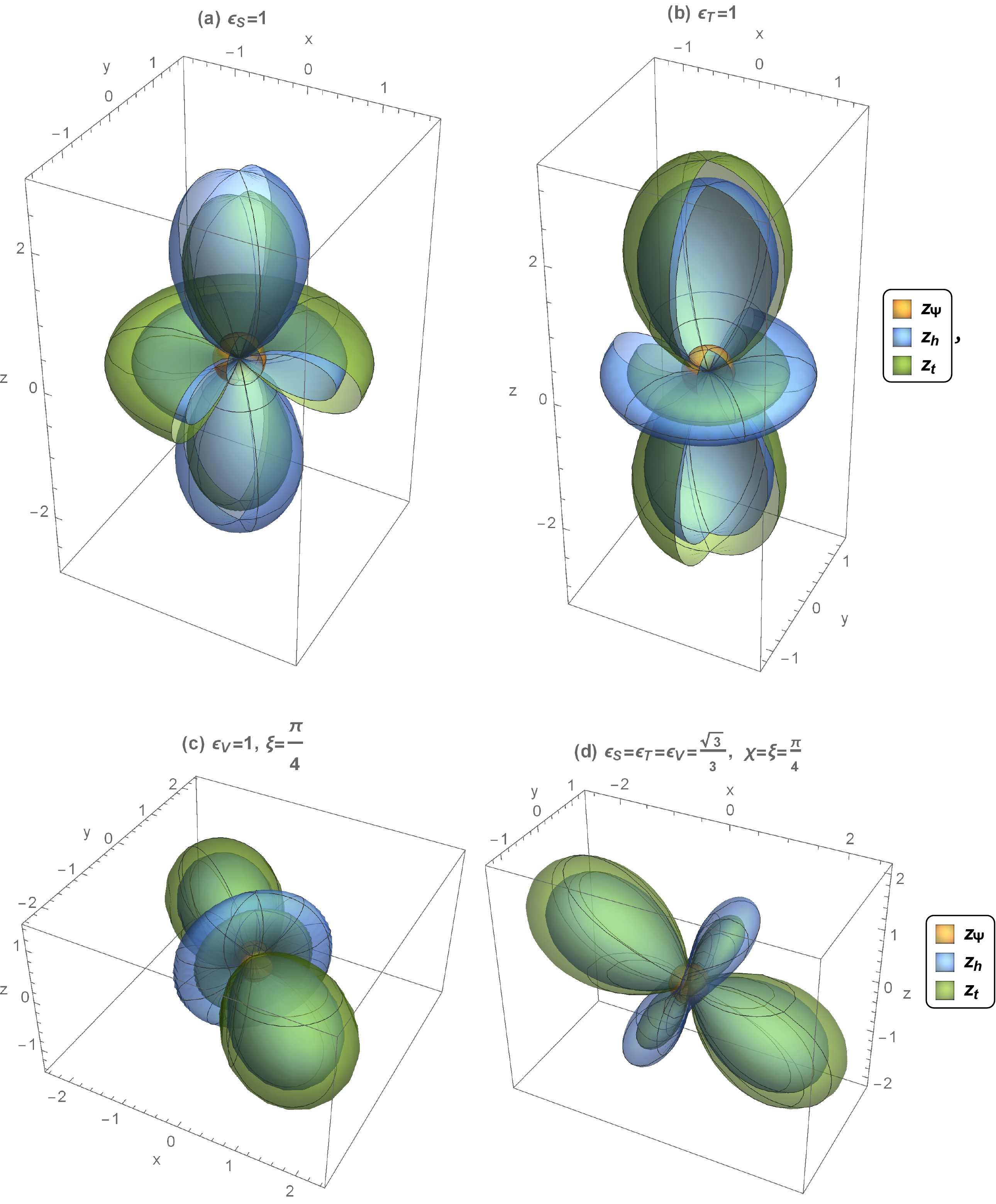}\caption{\label{redsft} The redshift induced by the tensor ULDM under four different cases: (a) only the scalar mode exists: $\epsS=1, \epsT=\epsV=0$; (b) only the tensor modes exist: $\epsT=1, \epsS=\epsV=0$; (c) only the vector modes exist: $\epsV=1, \epsS=\epsT=0$; (d) all five polarization modes exist with the specific values: $\epsV=\epsS=\epsT=\frac{\sqrt{3}}{3}$ and $\chi=\xi=\frac{\pi}{4}$. The orange surface and blue surface represent the contribution of the trace part \Eq{z_psi} and the traceless part \Eq{z_h}, respectively; their summation, the total redshift \Eq{z_t}, is depicted by the green surface.}
\end{figure}

Note that the total timing residual of the tensor ULDM depends on the amplitudes and azimuthal directions of all three sectors (scalar, vector, and tensor) as well as the pulsar's spatial orientation, resulting in a complicated anisotropy. \Fig{redsft} provides several examples of this complexity: subfigures (a)-(c) show scenarios where only one of the sectors exists, and subfigure (d) displays a more representative case containing all three sectors.


As reported in \cite{Kolb:2023dzp}, similar to the situation for spin-1 fields as discussed in \cite{Graham:2015rva},
the longitudinal scalar mode of the massive spin-2 field is more abundantly generated than the longitudinal-transverse vector and transverse tensor polarization modes.
Here and after, we restrict ourselves to the scalar mode of the tensor ULDM, namely case a) in \Fig{redsft} where $\epsS=1, \epsT=\epsV=0$. To highlight the physical implications behind this case, we will explicitly derive the timing residual induced by the scalar sector below. The corresponding equation of motion can be written as
\m
\begin{gathered}
M_{ij}=\frac{1}{\sqrt{6}}A(\bx)\cos(mt +\al(\bx))\begin{pmatrix}
-1& 0 & 0 \\
0 & -1 & 0\\
0 & 0 & 2
\end{pmatrix},
\end{gathered}
\n
and the nonvanishing energy-momentum components are
\m
T_{00}&=&\frac{1}{8}m^2 A(\bx)^2,\\
T_{xx}&=&T_{yy}=-\frac{1}{24}m^2 A(\bx)^2 \cos(2mt+2\al(\bx)),\\
T_{zz}&=&\frac{5}{24}m^2 A(\bx)^2 \cos(2mt+2\al(\bx)).
\n
In this case, the metric perturbations $\Psi(\bx)$ and $h_{ij}(\bx)$ induced by the trace and traceless parts of the energy-momentum can be expressed as
\begin{align}
\Psi(\bx)&=\Psi_0+\frac{1}{24}\pi G A(\bx)^2 \cos(2mt+2\alpha(\bx)), \notag\\
h_{ij}(\bx)&=\frac{1}{3}\pi G A(\bx)^2 \cos(2mt+2\alpha(\bx))
\begin{pmatrix}
1 & 0 &0  \\
0 & 1 & 0 \\
0 & 0 & -2
\end{pmatrix}.
\end{align}
These perturbations lead to redshifts $z_{\Psi}$ (\Eq{z_psi}) and $z_{h}$ (\Eq{z_h}) in the pulsar timing signal, which add up to yield the total redshift
\m
z_t &=&z_{\Psi}+z_{h} \notag\\
&=&-\frac{\pi G \rho_{\rm{TF}}}{3m^2}(1+6 \cos 2\theta)\times \[\cos(2mt+2\al(\bx_{e}))-\cos\(2m(t-|\bx_{p}|)+2\al(\bx_{p})\)\].
\n
Integrating the redshift results in the final timing residual,
\e\label{Rt_scalar}
R_{\rm{tensor}}=-\frac{\pi G \rho_{\rm{TF}}}{3m^3}(1+6 \cos2\theta)\sin (\alpha_{e}-\alpha_{p}) \cos \left(2 m t+\alpha_{e}+\alpha_{p}\right).
\q
The above expression can also be derived by simply plugging the conditions $\epsS=1$ and $\epsT=\epsV=0$ into \Eq{R_t}.



\begin{figure}[htbp!]
	\centering
	\includegraphics[width=0.3\textwidth]{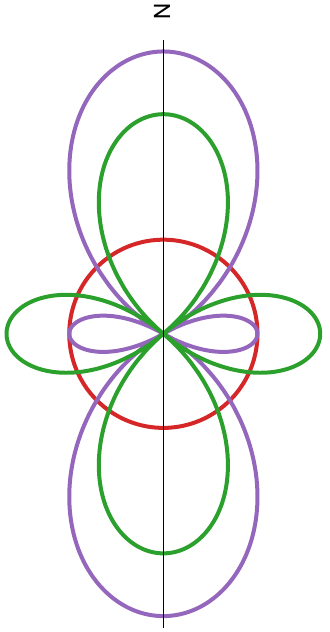}\caption{ \label{SVT}The redshift induced by oscillating ULDM with different spins. The red, purple and green lines represent the cases of the scalar, vector and tensor ULDM, respectively. The direction of oscillation is chosen along the z-axis for the vector and tensor ULDM.}
\end{figure}

Recall that the coherent oscillations of scalar and vector ULDM also generate timing residuals \citep{Khmelnitsky:2013lxt,Nomura:2019cvc},
\m
R_{\rm{scalar}}&=&\frac{\pi G \rho_{\rm{SF}}}{m^3}\sin (\alpha_{e}-\alpha_{p}) \cos \left(2 m t+\alpha_{e}+\alpha_{p}\right), \\
R_{\rm{vector}}&=&-\frac{\pi G \rho_{\rm{VF}}}{m^3}(1+2 \cos 2\theta)\sin (\alpha_{e}-\alpha_{p}) \cos \left(2 m t+\alpha_{e}+\alpha_{p}\right),
\n
where $\rho_{\rm{SF}}$ and $\rho_{\rm{VF}}$ denote the energy density of the scalar fields and vector fields, respectively.
As illustrated in \Fig{SVT}, the amplitude and angular dependence of the signals induced by ULDM of different spins differ significantly. Notably, the timing residuals of scalar ULDM are isotropic, while those of vector ULDM and tensor ULDM exhibit nontrivial direction dependence.

\section{Estimating the limits with current PTA}\label{sec4}

\begin{figure}[htbp!]
	\centering
	\includegraphics[width=0.8\textwidth]{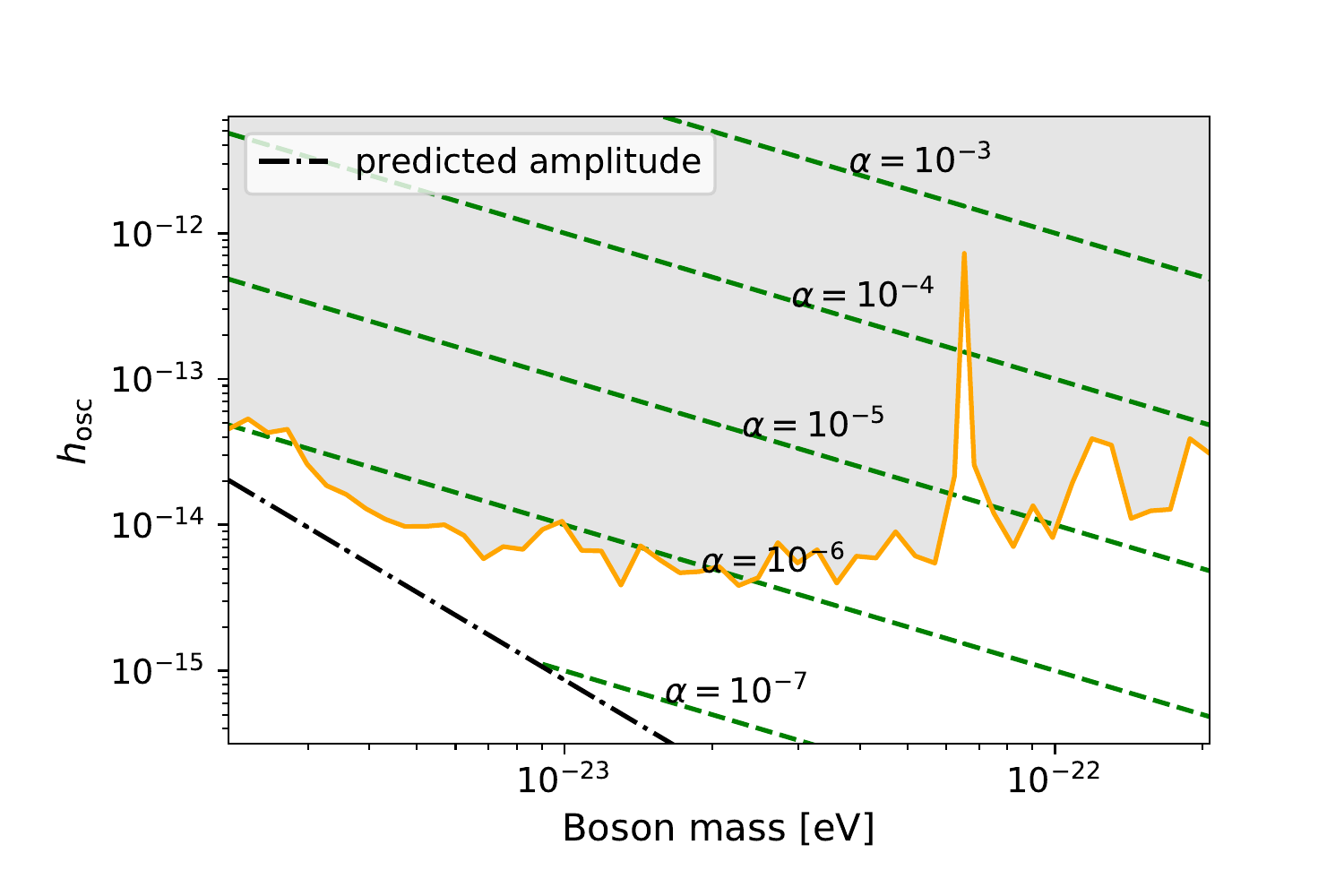}\caption{\label{h_osc} PPTA bounds on the potential amplitude $h_{\rm{osc}}$ induced by tensor ULDM versus the ULDM mass $m$. The black dot-dashed line denotes the model amplitude $h_{\rm{osc}}$ when assuming $\rho_{\rm{TF}}=0.4 \rm{GeV/cm^3}$. The green dashed line represents the deductive coupling effect from $h_{\rm{osc}}$ for several values of the ULDM coupling $\alpha$.  }
\end{figure}

Several PTA data sets have been utilized to search for the scalar or the vector ULDM \citep{Kato:2019bqz,Porayko:2018sfa,PPTA:2022eul}. To make an approximate prediction of the tensor ULDM using the existing results, we take the average of \Eq{R_t} over the celestial sphere $\hat{p}$, which gives
\m
\bar{R}_t=\sqrt{\frac{23}{15}}
\frac{\pi G \rho_{\rm{TF}}}{m^3}
\times \sin (\alpha_{e}-\alpha_{p}) \cos \left(2 m t+\alpha_{e}+\alpha_{p}\right).
\label{R_g1}
\n
Because of the system's symmetry, the average timing residuals from the gravitational effect do not depend on the five parameters describing the quadruple but only on the energy density of the tensor ULDM $\rho_{\rm{TF}}$. To utilize previous results and facilitate later discussion, we replace $\rho_{\rm{TF}}$ with the unique quantity $h_{\rm{osc}}$ from the gravitational effect that is directly constrained by PTA experiments.
Using the relationship \Eq{eq_h_osc}, we obtain
\m
\bar{R}_t=\frac{3}{4}\sqrt{\frac{23}{15}}
\frac{ h_{\rm{osc}}}{ m}
\times \sin (\alpha_{e}-\alpha_{p}) \cos \left(2 m t+\alpha_{e}+\alpha_{p}\right).
\label{R_g}
\n
In \Fig{h_osc}, we present the constraint on the amplitude $h_{\rm{osc}}$ for tensor ULDM, 
which is reproduced from the bounds on the vector ULDM by the second dataset release of the Parkes PTA collaboration (PPTA DR2) \citep{PPTA:2022eul}. 

Moreover, the spin-2 massive fields are universally coupled to the SM matter as a requirement of self-consistency of the model, which would cause a time delay for a pulse signal. The timing residual from the coupling effect is given by \citep{Armaleo:2020yml},
\m
\bar{R}_c=\sqrt{\frac{1}{15}}\frac{2\alpha \sqrt{\rho_{\rm{TF}}}}{ m^2 M_{\rm{P}}}\times \sin (\frac{\alpha_{e}-\alpha_{p}}{2}) \cos \left(m t+\frac{\alpha_{e}+\alpha_{p}}{2}\right),
\label{R_c}
\n
which depends on both the coupling $\alpha$ and the energy density $\rho_{\rm{TF}}$. 
Note that the timing residuals induced by the coupling effect, $\bar{R}_c$, and the gravitational effect, $\bar{R}_t$, have frequencies of $m$ and $2m$, respectively. If the spin-2 fields with a certain mass $m$ make up a significant fraction of the dark matter and the two effects are comparable, we will observe two signals in the PTA frequency band, with one having a frequency that is twice that of the other.
Theoretically, it is difficult to determine the values of the two quantities $\alpha$ and $\rho_{\rm{TF}}$ from the coupling effect, as they are degenerate. However, we can first use the pure gravitational effect, given by \Eq{R_g1}, which depends solely on $\rho_{\rm{TF}}$, to determine whether massive spin-2 fields make up all dark matter. If we cannot exclude this possibility, we can estimate an upper bound for $\alpha$ using \Eq{R_g} and \Eq{R_c},
\m
\alpha<\frac{3\sqrt{23}}{2}\frac{M_{\rm{P}}}{\sqrt{\rho_{\rm{0}}}}m h_{\rm{osc}}, 
\label{a_h}
\n
where $\rho_0=0.4 \rm{GeV/cm^3}$ is the measured local energy density \citep{Salucci:2010qr}. If the spin-2 fields cannot make up all the dark matter, we need to estimate $\alpha$ using the $\rho_{\rm{TF}}$ obtained from the gravitational effect.
In \Fig{h_osc}, we demonstrate several cases with different values of the coupling $\alpha$. We estimate that the coupling could get constrained at the level of $10^{-6}\sim 10^{-5}$ in the mass range $m<5\times 10^{-23} \rm{eV}$ with current PTA data. This result is consistent with the constraint derived in Ref.~\cite{Armaleo:2020yml} using the PPTA 12-year data set, because both results are obtained under the same assumption that $\rho_{\rm{TF}}=\rho_0$. However, we should emphasize that the green dashed line representing the coupling effect based on \Eq{a_h} cannot be extended to cross the left side of black dot-dashed line representing the predicted gravitational effect derived from $\rho_{0}$. Moreover, if future PTA experiments can constrain that $\rho_{\rm{TF}} < \rho_0$, then \Eq{a_h} is no longer valid and we need to reestimate the upper bound on $\alpha$ using an updated $\rho_{\rm{TF}}$.


\section{conclusion}\label{sec5}
In this work, we explore the gravitational effect of tensor ULDM on pulsar timing residuals. As a form of energy, the oscillating tensor field perturbs the spacetime geometry, resulting in the timing residuals that take the form of a sine function with a frequency equal to twice the particle mass, as shown in \Eq{R_t}. Although the timing residuals of tensor ULDM depend on all five parameters describing the spin-2 quadrupole, it is pointed by \citep{Kolb:2023dzp} that the gravitationally-produced massive spin-2 particles are predominantly longitudinally polarized, leading to a simpler expression \Eq{Rt_scalar} that still retains anisotropies. 
We find that the pulsar timing signal induced by the tensor ULDM exhibits a different angular dependence than that by the scalar and vector ULDM, as shown in \Fig{SVT}.

Besides the gravitational effect, the coupling between the tensor ULDM and SM matter also produces sine-shaped timing residuals, depending both on the coupling $\alpha$ and the energy density of tensor fields $\rho_{\rm{TF}}$. In previous studies, the upper bound of $\alpha$ was estimated by assuming that the tensor ULDM constitutes all dark matter, which is reasonable since the current sensitivity is insufficient to exclude the assumption. However, as experimental precision continues to improve, it may become possible to use the gravitational effects to determine whether massive spin-2 fields can account for all dark matter. In such cases, it would be necessary to re-evaluate the coupling parameter using an updated energy density of the tensor ULDM. We present an estimation using this approach with PPTA DR2 in \Fig{h_osc}.

While PTA's current sensitivity has limitations in efficiently constraining ULDM, many other cosmological and astrophysical probes have already provided important constraints on ULDM. For instance, cosmic microwave background put percent-level constraints on ULDM within the mass range $10^{-32}{\rm{eV}}<m<10^{-24}{\rm{eV}}$ \citep{Hlozek:2017zzf,Poulin:2018dzj}, and constraints from the galaxy ration curves \citep{Bar:2018acw} and Milky Way satellite population \citep{Nadler:2019zrb} impose a lower limit on the ULDM mass of $m \sim (10^{-22}{\rm{eV}}, 10^{-21}{\rm{eV}})$. The Lyman-alpha forest provides the most stringent constraint, disfavoring $m>2\times 10^{-20}{\rm{eV}}$ for ultralight fields constituting the entire dark matter \citep{Rogers:2020ltq}. 
Nevertheless, there is currently no consensus on ULDM mass constraints, with some exceptions favoring lower ULDM masses, such as a few times $10^{-23}{\rm{eV}}$ based on the stellar kinematics of dwarf spheroidal galaxies \citep{Gonzalez-Morales:2016yaf,Kendall:2019fep}.
On the one hand, many of these constraints are subject to their own astrophysical uncertainties. For instance, the constraint from Lyman-alpha forest relies on accurate modeling of fluctuations from temperature and impact of ionizing background, while the dwarf spheroidal bound may face challenges due to assumptions made about the sphericity of halos in these galaxies.
On the other hand, these constraints also largely depend on assumption regarding the structure formation in the ULDM scenario, and most astrophysical simulations are based on the scalar/axion setting. To provide more robust constraints, detailed interplay between ULDM physics and particle physics models, especially baryonic feedback, and simulations on higher-spin candidates capturing the polarization of vector and tensor ULDM are essential.
Furthermore, the possibility of ultralight fields not constituting all the dark matter or having multiple masses is also well-motivated. 
PTA experiments offer complementary tools to other experiments, independently constraining the energy density of ULDM. 
The forthcoming PTA project, square kilometer arrays (SKA), has the potential to detect a $10\%$ contribution of ULDM to the dark matter for $m<10^{-23}{\rm{eV}}$ within ten years and place interesting constraints on masses above $10^{-22}{\rm{eV}}$ if significantly high timing-precision is reached \citep{Porayko:2018sfa}. We believe cosmological and astrophysical probes across scales and epochs will offer a consistent understanding of the ULDM.

PTAs also hold the potential to distinguish different sources and even clarify the nature of the ultralight fields, since the timing residuals induced by scalar ULDM, vector ULDM, and tensor ULDM, as well as those induced by continuous gravitational waves emitted by a single supermassive black hole binary \cite{Zhu:2014rta,IPTA:2023ero}, exhibit distinct angular dependence. However, just as disentangling multiple astrophysical and cosmological stochastic gravitational wave backgrounds is challenging \citep{Kaiser:2022cma}, it may also be difficult to differentiate these monochromatic signals from each other. Therefore, investigating how they can be distinguished is an interesting research topic. One potential solution is to study the correlation pattern of different signals \citep{Unal:2022ooa}. The underlying concept is that when the same common signal modulates the arrival time of pulses from various pulsars, the resulting timing residuals of any two pulsars will be correlated. Just as gravitational waves generate a Hellings-Downs pattern \citep{Hellings:1983fr} of correlation, the ULDM signal with different spins should exhibit distinct correlation forms. While some studies have pointed out the existence of the correlations and attempted to calculate the form in specific cases \citep{Armaleo:2020yml,Liang:2021bct}, further in-depth research is still needed.

\begin{acknowledgments}
We acknowledge the use of HPC Cluster of ITP-CAS. QGH is supported by the grants from NSFC (Grant No.~12250010, 11975019, 11991052, 12047503), Key Research Program of Frontier Sciences, CAS, Grant No.~ZDBS-LY-7009, CAS Project for Young Scientists in Basic Research YSBR-006, the Key Research Program of the Chinese Academy of Sciences (Grant No.~XDPB15). 
ZCC is supported by the National Natural Science Foundation of China (Grant No.~12247176 and No.~12247112) and the China Postdoctoral Science Foundation Fellowship No.~2022M710429.
\end{acknowledgments}
\bibliography{utdm.bib}

\begin{thebibliography}{64}%
\makeatletter
\providecommand \@ifxundefined [1]{%
 \@ifx{#1\undefined}
}%
\providecommand \@ifnum [1]{%
 \ifnum #1\expandafter \@firstoftwo
 \else \expandafter \@secondoftwo
 \fi
}%
\providecommand \@ifx [1]{%
 \ifx #1\expandafter \@firstoftwo
 \else \expandafter \@secondoftwo
 \fi
}%
\providecommand \natexlab [1]{#1}%
\providecommand \enquote  [1]{``#1''}%
\providecommand \bibnamefont  [1]{#1}%
\providecommand \bibfnamefont [1]{#1}%
\providecommand \citenamefont [1]{#1}%
\providecommand \href@noop [0]{\@secondoftwo}%
\providecommand \href [0]{\begingroup \@sanitize@url \@href}%
\providecommand \@href[1]{\@@startlink{#1}\@@href}%
\providecommand \@@href[1]{\endgroup#1\@@endlink}%
\providecommand \@sanitize@url [0]{\catcode `\\12\catcode `\$12\catcode
  `\&12\catcode `\#12\catcode `\^12\catcode `\_12\catcode `\%12\relax}%
\providecommand \@@startlink[1]{}%
\providecommand \@@endlink[0]{}%
\providecommand \url  [0]{\begingroup\@sanitize@url \@url }%
\providecommand \@url [1]{\endgroup\@href {#1}{\urlprefix }}%
\providecommand \urlprefix  [0]{URL }%
\providecommand \Eprint [0]{\href }%
\providecommand \doibase [0]{http://dx.doi.org/}%
\providecommand \selectlanguage [0]{\@gobble}%
\providecommand \bibinfo  [0]{\@secondoftwo}%
\providecommand \bibfield  [0]{\@secondoftwo}%
\providecommand \translation [1]{[#1]}%
\providecommand \BibitemOpen [0]{}%
\providecommand \bibitemStop [0]{}%
\providecommand \bibitemNoStop [0]{.\EOS\space}%
\providecommand \EOS [0]{\spacefactor3000\relax}%
\providecommand \BibitemShut  [1]{\csname bibitem#1\endcsname}%
\let\auto@bib@innerbib\@empty
\bibitem [{\citenamefont {{Rubin}}\ \emph {et~al.}(1980)\citenamefont
  {{Rubin}}, \citenamefont {{Ford}},\ and\ \citenamefont
  {{Thonnard}}}]{1980ApJ...238..471R}%
  \BibitemOpen
  \bibfield  {author} {\bibinfo {author} {\bibfnamefont {V.~C.}\ \bibnamefont
  {{Rubin}}}, \bibinfo {author} {\bibfnamefont {Jr.}\ \bibnamefont {{Ford}},
  \bibfnamefont {W.~K.}}, \ and\ \bibinfo {author} {\bibfnamefont
  {N.}~\bibnamefont {{Thonnard}}},\ }\bibfield  {title} {\enquote {\bibinfo
  {title} {{Rotational properties of 21 SC galaxies with a large range of
  luminosities and radii, from NGC 4605 (R=4kpc) to UGC 2885 (R=122kpc).}}}\
  }\href {\doibase 10.1086/158003} {\bibfield  {journal} {\bibinfo  {journal}
  {\apj}\ }\textbf {\bibinfo {volume} {238}},\ \bibinfo {pages} {471--487}
  (\bibinfo {year} {1980})}\BibitemShut {NoStop}%
\bibitem [{\citenamefont {{Rubin}}\ \emph {et~al.}(1982)\citenamefont
  {{Rubin}}, \citenamefont {{Ford}}, \citenamefont {{Thonnard}},\ and\
  \citenamefont {{Burstein}}}]{1982ApJ...261..439R}%
  \BibitemOpen
  \bibfield  {author} {\bibinfo {author} {\bibfnamefont {V.~C.}\ \bibnamefont
  {{Rubin}}}, \bibinfo {author} {\bibfnamefont {Jr.}\ \bibnamefont {{Ford}},
  \bibfnamefont {W.~K.}}, \bibinfo {author} {\bibfnamefont {N.}~\bibnamefont
  {{Thonnard}}}, \ and\ \bibinfo {author} {\bibfnamefont {D.}~\bibnamefont
  {{Burstein}}},\ }\bibfield  {title} {\enquote {\bibinfo {title} {{Rotational
  properties of 23Sb galaxies.}}}\ }\href {\doibase 10.1086/160355} {\bibfield
  {journal} {\bibinfo  {journal} {\apj}\ }\textbf {\bibinfo {volume} {261}},\
  \bibinfo {pages} {439--456} (\bibinfo {year} {1982})}\BibitemShut {NoStop}%
\bibitem [{\citenamefont {{Faber}}\ and\ \citenamefont
  {{Jackson}}(1976)}]{1976ApJ...204..668F}%
  \BibitemOpen
  \bibfield  {author} {\bibinfo {author} {\bibfnamefont {S.~M.}\ \bibnamefont
  {{Faber}}}\ and\ \bibinfo {author} {\bibfnamefont {R.~E.}\ \bibnamefont
  {{Jackson}}},\ }\bibfield  {title} {\enquote {\bibinfo {title} {{Velocity
  dispersions and mass-to-light ratios for elliptical galaxies.}}}\ }\href
  {\doibase 10.1086/154215} {\bibfield  {journal} {\bibinfo  {journal} {\apj}\
  }\textbf {\bibinfo {volume} {204}},\ \bibinfo {pages} {668--683} (\bibinfo
  {year} {1976})}\BibitemShut {NoStop}%
\bibitem [{\citenamefont {Massey}\ \emph {et~al.}(2010)\citenamefont {Massey},
  \citenamefont {Kitching},\ and\ \citenamefont {Richard}}]{Massey:2010hh}%
  \BibitemOpen
  \bibfield  {author} {\bibinfo {author} {\bibfnamefont {Richard}\ \bibnamefont
  {Massey}}, \bibinfo {author} {\bibfnamefont {Thomas}\ \bibnamefont
  {Kitching}}, \ and\ \bibinfo {author} {\bibfnamefont {Johan}\ \bibnamefont
  {Richard}},\ }\bibfield  {title} {\enquote {\bibinfo {title} {{The dark
  matter of gravitational lensing}},}\ }\href {\doibase
  10.1088/0034-4885/73/8/086901} {\bibfield  {journal} {\bibinfo  {journal}
  {Rept. Prog. Phys.}\ }\textbf {\bibinfo {volume} {73}},\ \bibinfo {pages}
  {086901} (\bibinfo {year} {2010})},\ \Eprint {http://arxiv.org/abs/1001.1739}
  {arXiv:1001.1739 [astro-ph.CO]} \BibitemShut {NoStop}%
\bibitem [{\citenamefont {Aghanim}\ \emph {et~al.}(2020)\citenamefont {Aghanim}
  \emph {et~al.}}]{Planck:2018vyg}%
  \BibitemOpen
  \bibfield  {author} {\bibinfo {author} {\bibfnamefont {N.}~\bibnamefont
  {Aghanim}} \emph {et~al.} (\bibinfo {collaboration} {Planck}),\ }\bibfield
  {title} {\enquote {\bibinfo {title} {{Planck 2018 results. VI. Cosmological
  parameters}},}\ }\href {\doibase 10.1051/0004-6361/201833910} {\bibfield
  {journal} {\bibinfo  {journal} {Astron. Astrophys.}\ }\textbf {\bibinfo
  {volume} {641}},\ \bibinfo {pages} {A6} (\bibinfo {year} {2020})},\ \bibinfo
  {note} {[Erratum: Astron.Astrophys. 652, C4 (2021)]},\ \Eprint
  {http://arxiv.org/abs/1807.06209} {arXiv:1807.06209 [astro-ph.CO]}
  \BibitemShut {NoStop}%
\bibitem [{\citenamefont {Bertone}\ \emph {et~al.}(2005)\citenamefont
  {Bertone}, \citenamefont {Hooper},\ and\ \citenamefont
  {Silk}}]{Bertone:2004pz}%
  \BibitemOpen
  \bibfield  {author} {\bibinfo {author} {\bibfnamefont {Gianfranco}\
  \bibnamefont {Bertone}}, \bibinfo {author} {\bibfnamefont {Dan}\ \bibnamefont
  {Hooper}}, \ and\ \bibinfo {author} {\bibfnamefont {Joseph}\ \bibnamefont
  {Silk}},\ }\bibfield  {title} {\enquote {\bibinfo {title} {{Particle dark
  matter: Evidence, candidates and constraints}},}\ }\href {\doibase
  10.1016/j.physrep.2004.08.031} {\bibfield  {journal} {\bibinfo  {journal}
  {Phys. Rept.}\ }\textbf {\bibinfo {volume} {405}},\ \bibinfo {pages}
  {279--390} (\bibinfo {year} {2005})},\ \Eprint
  {http://arxiv.org/abs/hep-ph/0404175} {arXiv:hep-ph/0404175} \BibitemShut
  {NoStop}%
\bibitem [{\citenamefont {Feng}(2010)}]{Feng:2010gw}%
  \BibitemOpen
  \bibfield  {author} {\bibinfo {author} {\bibfnamefont {Jonathan~L.}\
  \bibnamefont {Feng}},\ }\bibfield  {title} {\enquote {\bibinfo {title} {{Dark
  Matter Candidates from Particle Physics and Methods of Detection}},}\ }\href
  {\doibase 10.1146/annurev-astro-082708-101659} {\bibfield  {journal}
  {\bibinfo  {journal} {Ann. Rev. Astron. Astrophys.}\ }\textbf {\bibinfo
  {volume} {48}},\ \bibinfo {pages} {495--545} (\bibinfo {year} {2010})},\
  \Eprint {http://arxiv.org/abs/1003.0904} {arXiv:1003.0904 [astro-ph.CO]}
  \BibitemShut {NoStop}%
\bibitem [{\citenamefont {Bertone}\ and\ \citenamefont
  {Tait}(2018)}]{Bertone:2018krk}%
  \BibitemOpen
  \bibfield  {author} {\bibinfo {author} {\bibfnamefont {Gianfranco}\
  \bibnamefont {Bertone}}\ and\ \bibinfo {author} {\bibfnamefont {Tim}\
  \bibnamefont {Tait}, \bibfnamefont {M.~P.}},\ }\bibfield  {title} {\enquote
  {\bibinfo {title} {{A new era in the search for dark matter}},}\ }\href
  {\doibase 10.1038/s41586-018-0542-z} {\bibfield  {journal} {\bibinfo
  {journal} {Nature}\ }\textbf {\bibinfo {volume} {562}},\ \bibinfo {pages}
  {51--56} (\bibinfo {year} {2018})},\ \Eprint
  {http://arxiv.org/abs/1810.01668} {arXiv:1810.01668 [astro-ph.CO]}
  \BibitemShut {NoStop}%
\bibitem [{\citenamefont {Marsh}(2016)}]{Marsh:2015xka}%
  \BibitemOpen
  \bibfield  {author} {\bibinfo {author} {\bibfnamefont {David J.~E.}\
  \bibnamefont {Marsh}},\ }\bibfield  {title} {\enquote {\bibinfo {title}
  {{Axion Cosmology}},}\ }\href {\doibase 10.1016/j.physrep.2016.06.005}
  {\bibfield  {journal} {\bibinfo  {journal} {Phys. Rept.}\ }\textbf {\bibinfo
  {volume} {643}},\ \bibinfo {pages} {1--79} (\bibinfo {year} {2016})},\
  \Eprint {http://arxiv.org/abs/1510.07633} {arXiv:1510.07633 [astro-ph.CO]}
  \BibitemShut {NoStop}%
\bibitem [{\citenamefont {Alcock}\ \emph {et~al.}(2000)\citenamefont {Alcock}
  \emph {et~al.}}]{MACHO:2000qbb}%
  \BibitemOpen
  \bibfield  {author} {\bibinfo {author} {\bibfnamefont {C.}~\bibnamefont
  {Alcock}} \emph {et~al.} (\bibinfo {collaboration} {MACHO}),\ }\bibfield
  {title} {\enquote {\bibinfo {title} {{The MACHO project: Microlensing results
  from 5.7 years of LMC observations}},}\ }\href {\doibase 10.1086/309512}
  {\bibfield  {journal} {\bibinfo  {journal} {Astrophys. J.}\ }\textbf
  {\bibinfo {volume} {542}},\ \bibinfo {pages} {281--307} (\bibinfo {year}
  {2000})},\ \Eprint {http://arxiv.org/abs/astro-ph/0001272}
  {arXiv:astro-ph/0001272} \BibitemShut {NoStop}%
\bibitem [{\citenamefont {Schumann}(2019)}]{Schumann:2019eaa}%
  \BibitemOpen
  \bibfield  {author} {\bibinfo {author} {\bibfnamefont {Marc}\ \bibnamefont
  {Schumann}},\ }\bibfield  {title} {\enquote {\bibinfo {title} {{Direct
  Detection of WIMP Dark Matter: Concepts and Status}},}\ }\href {\doibase
  10.1088/1361-6471/ab2ea5} {\bibfield  {journal} {\bibinfo  {journal} {J.
  Phys. G}\ }\textbf {\bibinfo {volume} {46}},\ \bibinfo {pages} {103003}
  (\bibinfo {year} {2019})},\ \Eprint {http://arxiv.org/abs/1903.03026}
  {arXiv:1903.03026 [astro-ph.CO]} \BibitemShut {NoStop}%
\bibitem [{\citenamefont {Hu}\ \emph {et~al.}(2000)\citenamefont {Hu},
  \citenamefont {Barkana},\ and\ \citenamefont {Gruzinov}}]{Hu:2000ke}%
  \BibitemOpen
  \bibfield  {author} {\bibinfo {author} {\bibfnamefont {Wayne}\ \bibnamefont
  {Hu}}, \bibinfo {author} {\bibfnamefont {Rennan}\ \bibnamefont {Barkana}}, \
  and\ \bibinfo {author} {\bibfnamefont {Andrei}\ \bibnamefont {Gruzinov}},\
  }\bibfield  {title} {\enquote {\bibinfo {title} {{Cold and fuzzy dark
  matter}},}\ }\href {\doibase 10.1103/PhysRevLett.85.1158} {\bibfield
  {journal} {\bibinfo  {journal} {Phys. Rev. Lett.}\ }\textbf {\bibinfo
  {volume} {85}},\ \bibinfo {pages} {1158--1161} (\bibinfo {year} {2000})},\
  \Eprint {http://arxiv.org/abs/astro-ph/0003365} {arXiv:astro-ph/0003365}
  \BibitemShut {NoStop}%
\bibitem [{\citenamefont {Hui}\ \emph {et~al.}(2017)\citenamefont {Hui},
  \citenamefont {Ostriker}, \citenamefont {Tremaine},\ and\ \citenamefont
  {Witten}}]{Hui:2016ltb}%
  \BibitemOpen
  \bibfield  {author} {\bibinfo {author} {\bibfnamefont {Lam}\ \bibnamefont
  {Hui}}, \bibinfo {author} {\bibfnamefont {Jeremiah~P.}\ \bibnamefont
  {Ostriker}}, \bibinfo {author} {\bibfnamefont {Scott}\ \bibnamefont
  {Tremaine}}, \ and\ \bibinfo {author} {\bibfnamefont {Edward}\ \bibnamefont
  {Witten}},\ }\bibfield  {title} {\enquote {\bibinfo {title} {{Ultralight
  scalars as cosmological dark matter}},}\ }\href {\doibase
  10.1103/PhysRevD.95.043541} {\bibfield  {journal} {\bibinfo  {journal} {Phys.
  Rev. D}\ }\textbf {\bibinfo {volume} {95}},\ \bibinfo {pages} {043541}
  (\bibinfo {year} {2017})},\ \Eprint {http://arxiv.org/abs/1610.08297}
  {arXiv:1610.08297 [astro-ph.CO]} \BibitemShut {NoStop}%
\bibitem [{\citenamefont {Fox}\ \emph {et~al.}(2004)\citenamefont {Fox},
  \citenamefont {Pierce},\ and\ \citenamefont {Thomas}}]{Fox:2004kb}%
  \BibitemOpen
  \bibfield  {author} {\bibinfo {author} {\bibfnamefont {Patrick}\ \bibnamefont
  {Fox}}, \bibinfo {author} {\bibfnamefont {Aaron}\ \bibnamefont {Pierce}}, \
  and\ \bibinfo {author} {\bibfnamefont {Scott~D.}\ \bibnamefont {Thomas}},\
  }\bibfield  {title} {\enquote {\bibinfo {title} {{Probing a QCD string axion
  with precision cosmological measurements}},}\ }\href@noop {} {\  (\bibinfo
  {year} {2004})},\ \Eprint {http://arxiv.org/abs/hep-th/0409059}
  {arXiv:hep-th/0409059} \BibitemShut {NoStop}%
\bibitem [{\citenamefont {Niemeyer}(2019)}]{Niemeyer:2019aqm}%
  \BibitemOpen
  \bibfield  {author} {\bibinfo {author} {\bibfnamefont {Jens~C.}\ \bibnamefont
  {Niemeyer}},\ }\bibfield  {title} {\enquote {\bibinfo {title} {{Small-scale
  structure of fuzzy and axion-like dark matter}},}\ }\href {\doibase
  10.1016/j.ppnp.2020.103787} {\bibfield  {journal} {\bibinfo  {journal} {Prog.
  Part. Nucl. Phys.}\ } (\bibinfo {year} {2019}),\
  10.1016/j.ppnp.2020.103787},\ \Eprint {http://arxiv.org/abs/1912.07064}
  {arXiv:1912.07064 [astro-ph.CO]} \BibitemShut {NoStop}%
\bibitem [{\citenamefont {Gentile}\ \emph {et~al.}(2004)\citenamefont
  {Gentile}, \citenamefont {Salucci}, \citenamefont {Klein}, \citenamefont
  {Vergani},\ and\ \citenamefont {Kalberla}}]{Gentile:2004tb}%
  \BibitemOpen
  \bibfield  {author} {\bibinfo {author} {\bibfnamefont {Gianfranco}\
  \bibnamefont {Gentile}}, \bibinfo {author} {\bibfnamefont {P.}~\bibnamefont
  {Salucci}}, \bibinfo {author} {\bibfnamefont {U.}~\bibnamefont {Klein}},
  \bibinfo {author} {\bibfnamefont {D.}~\bibnamefont {Vergani}}, \ and\
  \bibinfo {author} {\bibfnamefont {P.}~\bibnamefont {Kalberla}},\ }\bibfield
  {title} {\enquote {\bibinfo {title} {{The Cored distribution of dark matter
  in spiral galaxies}},}\ }\href {\doibase 10.1111/j.1365-2966.2004.07836.x}
  {\bibfield  {journal} {\bibinfo  {journal} {Mon. Not. Roy. Astron. Soc.}\
  }\textbf {\bibinfo {volume} {351}},\ \bibinfo {pages} {903} (\bibinfo {year}
  {2004})},\ \Eprint {http://arxiv.org/abs/astro-ph/0403154}
  {arXiv:astro-ph/0403154} \BibitemShut {NoStop}%
\bibitem [{\citenamefont {de~Blok}(2010)}]{deBlok:2009sp}%
  \BibitemOpen
  \bibfield  {author} {\bibinfo {author} {\bibfnamefont {W.~J.~G.}\
  \bibnamefont {de~Blok}},\ }\bibfield  {title} {\enquote {\bibinfo {title}
  {{The Core-Cusp Problem}},}\ }\href {\doibase 10.1155/2010/789293} {\bibfield
   {journal} {\bibinfo  {journal} {Adv. Astron.}\ }\textbf {\bibinfo {volume}
  {2010}},\ \bibinfo {pages} {789293} (\bibinfo {year} {2010})},\ \Eprint
  {http://arxiv.org/abs/0910.3538} {arXiv:0910.3538 [astro-ph.CO]} \BibitemShut
  {NoStop}%
\bibitem [{\citenamefont {Moore}\ \emph {et~al.}(1999)\citenamefont {Moore},
  \citenamefont {Ghigna}, \citenamefont {Governato}, \citenamefont {Lake},
  \citenamefont {Quinn}, \citenamefont {Stadel},\ and\ \citenamefont
  {Tozzi}}]{Moore:1999nt}%
  \BibitemOpen
  \bibfield  {author} {\bibinfo {author} {\bibfnamefont {B.}~\bibnamefont
  {Moore}}, \bibinfo {author} {\bibfnamefont {S.}~\bibnamefont {Ghigna}},
  \bibinfo {author} {\bibfnamefont {F.}~\bibnamefont {Governato}}, \bibinfo
  {author} {\bibfnamefont {G.}~\bibnamefont {Lake}}, \bibinfo {author}
  {\bibfnamefont {Thomas~R.}\ \bibnamefont {Quinn}}, \bibinfo {author}
  {\bibfnamefont {J.}~\bibnamefont {Stadel}}, \ and\ \bibinfo {author}
  {\bibfnamefont {P.}~\bibnamefont {Tozzi}},\ }\bibfield  {title} {\enquote
  {\bibinfo {title} {{Dark matter substructure within galactic halos}},}\
  }\href {\doibase 10.1086/312287} {\bibfield  {journal} {\bibinfo  {journal}
  {Astrophys. J. Lett.}\ }\textbf {\bibinfo {volume} {524}},\ \bibinfo {pages}
  {L19--L22} (\bibinfo {year} {1999})},\ \Eprint
  {http://arxiv.org/abs/astro-ph/9907411} {arXiv:astro-ph/9907411} \BibitemShut
  {NoStop}%
\bibitem [{\citenamefont {Klypin}\ \emph {et~al.}(1999)\citenamefont {Klypin},
  \citenamefont {Kravtsov}, \citenamefont {Valenzuela},\ and\ \citenamefont
  {Prada}}]{Klypin:1999uc}%
  \BibitemOpen
  \bibfield  {author} {\bibinfo {author} {\bibfnamefont {Anatoly~A.}\
  \bibnamefont {Klypin}}, \bibinfo {author} {\bibfnamefont {Andrey~V.}\
  \bibnamefont {Kravtsov}}, \bibinfo {author} {\bibfnamefont {Octavio}\
  \bibnamefont {Valenzuela}}, \ and\ \bibinfo {author} {\bibfnamefont
  {Francisco}\ \bibnamefont {Prada}},\ }\bibfield  {title} {\enquote {\bibinfo
  {title} {{Where are the missing Galactic satellites?}}}\ }\href {\doibase
  10.1086/307643} {\bibfield  {journal} {\bibinfo  {journal} {Astrophys. J.}\
  }\textbf {\bibinfo {volume} {522}},\ \bibinfo {pages} {82--92} (\bibinfo
  {year} {1999})},\ \Eprint {http://arxiv.org/abs/astro-ph/9901240}
  {arXiv:astro-ph/9901240} \BibitemShut {NoStop}%
\bibitem [{\citenamefont {Khmelnitsky}\ and\ \citenamefont
  {Rubakov}(2014)}]{Khmelnitsky:2013lxt}%
  \BibitemOpen
  \bibfield  {author} {\bibinfo {author} {\bibfnamefont {Andrei}\ \bibnamefont
  {Khmelnitsky}}\ and\ \bibinfo {author} {\bibfnamefont {Valery}\ \bibnamefont
  {Rubakov}},\ }\bibfield  {title} {\enquote {\bibinfo {title} {{Pulsar timing
  signal from ultralight scalar dark matter}},}\ }\href {\doibase
  10.1088/1475-7516/2014/02/019} {\bibfield  {journal} {\bibinfo  {journal}
  {JCAP}\ }\textbf {\bibinfo {volume} {02}},\ \bibinfo {pages} {019} (\bibinfo
  {year} {2014})},\ \Eprint {http://arxiv.org/abs/1309.5888} {arXiv:1309.5888
  [astro-ph.CO]} \BibitemShut {NoStop}%
\bibitem [{\citenamefont {Arvanitaki}\ \emph {et~al.}(2010)\citenamefont
  {Arvanitaki}, \citenamefont {Dimopoulos}, \citenamefont {Dubovsky},
  \citenamefont {Kaloper},\ and\ \citenamefont
  {March-Russell}}]{Arvanitaki:2009fg}%
  \BibitemOpen
  \bibfield  {author} {\bibinfo {author} {\bibfnamefont {Asimina}\ \bibnamefont
  {Arvanitaki}}, \bibinfo {author} {\bibfnamefont {Savas}\ \bibnamefont
  {Dimopoulos}}, \bibinfo {author} {\bibfnamefont {Sergei}\ \bibnamefont
  {Dubovsky}}, \bibinfo {author} {\bibfnamefont {Nemanja}\ \bibnamefont
  {Kaloper}}, \ and\ \bibinfo {author} {\bibfnamefont {John}\ \bibnamefont
  {March-Russell}},\ }\bibfield  {title} {\enquote {\bibinfo {title} {{String
  Axiverse}},}\ }\href {\doibase 10.1103/PhysRevD.81.123530} {\bibfield
  {journal} {\bibinfo  {journal} {Phys. Rev. D}\ }\textbf {\bibinfo {volume}
  {81}},\ \bibinfo {pages} {123530} (\bibinfo {year} {2010})},\ \Eprint
  {http://arxiv.org/abs/0905.4720} {arXiv:0905.4720 [hep-th]} \BibitemShut
  {NoStop}%
\bibitem [{\citenamefont {Hlozek}\ \emph {et~al.}(2015)\citenamefont {Hlozek},
  \citenamefont {Grin}, \citenamefont {Marsh},\ and\ \citenamefont
  {Ferreira}}]{Hlozek:2014lca}%
  \BibitemOpen
  \bibfield  {author} {\bibinfo {author} {\bibfnamefont {Ren\'ee}\ \bibnamefont
  {Hlozek}}, \bibinfo {author} {\bibfnamefont {Daniel}\ \bibnamefont {Grin}},
  \bibinfo {author} {\bibfnamefont {David J.~E.}\ \bibnamefont {Marsh}}, \ and\
  \bibinfo {author} {\bibfnamefont {Pedro~G.}\ \bibnamefont {Ferreira}},\
  }\bibfield  {title} {\enquote {\bibinfo {title} {{A search for ultralight
  axions using precision cosmological data}},}\ }\href {\doibase
  10.1103/PhysRevD.91.103512} {\bibfield  {journal} {\bibinfo  {journal} {Phys.
  Rev. D}\ }\textbf {\bibinfo {volume} {91}},\ \bibinfo {pages} {103512}
  (\bibinfo {year} {2015})},\ \Eprint {http://arxiv.org/abs/1410.2896}
  {arXiv:1410.2896 [astro-ph.CO]} \BibitemShut {NoStop}%
\bibitem [{\citenamefont {Payez}\ \emph {et~al.}(2015)\citenamefont {Payez},
  \citenamefont {Evoli}, \citenamefont {Fischer}, \citenamefont {Giannotti},
  \citenamefont {Mirizzi},\ and\ \citenamefont {Ringwald}}]{Payez:2014xsa}%
  \BibitemOpen
  \bibfield  {author} {\bibinfo {author} {\bibfnamefont {Alexandre}\
  \bibnamefont {Payez}}, \bibinfo {author} {\bibfnamefont {Carmelo}\
  \bibnamefont {Evoli}}, \bibinfo {author} {\bibfnamefont {Tobias}\
  \bibnamefont {Fischer}}, \bibinfo {author} {\bibfnamefont {Maurizio}\
  \bibnamefont {Giannotti}}, \bibinfo {author} {\bibfnamefont {Alessandro}\
  \bibnamefont {Mirizzi}}, \ and\ \bibinfo {author} {\bibfnamefont {Andreas}\
  \bibnamefont {Ringwald}},\ }\bibfield  {title} {\enquote {\bibinfo {title}
  {{Revisiting the SN1987A gamma-ray limit on ultralight axion-like
  particles}},}\ }\href {\doibase 10.1088/1475-7516/2015/02/006} {\bibfield
  {journal} {\bibinfo  {journal} {JCAP}\ }\textbf {\bibinfo {volume} {02}},\
  \bibinfo {pages} {006} (\bibinfo {year} {2015})},\ \Eprint
  {http://arxiv.org/abs/1410.3747} {arXiv:1410.3747 [astro-ph.HE]} \BibitemShut
  {NoStop}%
\bibitem [{\citenamefont {Ivanov}\ \emph {et~al.}(2019)\citenamefont {Ivanov},
  \citenamefont {Kovalev}, \citenamefont {Lister}, \citenamefont {Panin},
  \citenamefont {Pushkarev}, \citenamefont {Savolainen},\ and\ \citenamefont
  {Troitsky}}]{Ivanov:2018byi}%
  \BibitemOpen
  \bibfield  {author} {\bibinfo {author} {\bibfnamefont {M.~M.}\ \bibnamefont
  {Ivanov}}, \bibinfo {author} {\bibfnamefont {Y.~Y.}\ \bibnamefont {Kovalev}},
  \bibinfo {author} {\bibfnamefont {M.~L.}\ \bibnamefont {Lister}}, \bibinfo
  {author} {\bibfnamefont {A.~G.}\ \bibnamefont {Panin}}, \bibinfo {author}
  {\bibfnamefont {A.~B.}\ \bibnamefont {Pushkarev}}, \bibinfo {author}
  {\bibfnamefont {T.}~\bibnamefont {Savolainen}}, \ and\ \bibinfo {author}
  {\bibfnamefont {S.~V.}\ \bibnamefont {Troitsky}},\ }\bibfield  {title}
  {\enquote {\bibinfo {title} {{Constraining the photon coupling of ultra-light
  dark-matter axion-like particles by polarization variations of parsec-scale
  jets in active galaxies}},}\ }\href {\doibase 10.1088/1475-7516/2019/02/059}
  {\bibfield  {journal} {\bibinfo  {journal} {JCAP}\ }\textbf {\bibinfo
  {volume} {02}},\ \bibinfo {pages} {059} (\bibinfo {year} {2019})},\ \Eprint
  {http://arxiv.org/abs/1811.10997} {arXiv:1811.10997 [astro-ph.CO]}
  \BibitemShut {NoStop}%
\bibitem [{\citenamefont {Pierce}\ \emph {et~al.}(2018)\citenamefont {Pierce},
  \citenamefont {Riles},\ and\ \citenamefont {Zhao}}]{Pierce:2018xmy}%
  \BibitemOpen
  \bibfield  {author} {\bibinfo {author} {\bibfnamefont {Aaron}\ \bibnamefont
  {Pierce}}, \bibinfo {author} {\bibfnamefont {Keith}\ \bibnamefont {Riles}}, \
  and\ \bibinfo {author} {\bibfnamefont {Yue}\ \bibnamefont {Zhao}},\
  }\bibfield  {title} {\enquote {\bibinfo {title} {{Searching for Dark Photon
  Dark Matter with Gravitational Wave Detectors}},}\ }\href {\doibase
  10.1103/PhysRevLett.121.061102} {\bibfield  {journal} {\bibinfo  {journal}
  {Phys. Rev. Lett.}\ }\textbf {\bibinfo {volume} {121}},\ \bibinfo {pages}
  {061102} (\bibinfo {year} {2018})},\ \Eprint
  {http://arxiv.org/abs/1801.10161} {arXiv:1801.10161 [hep-ph]} \BibitemShut
  {NoStop}%
\bibitem [{\citenamefont {Nomura}\ \emph {et~al.}(2020)\citenamefont {Nomura},
  \citenamefont {Ito},\ and\ \citenamefont {Soda}}]{Nomura:2019cvc}%
  \BibitemOpen
  \bibfield  {author} {\bibinfo {author} {\bibfnamefont {Kimihiro}\
  \bibnamefont {Nomura}}, \bibinfo {author} {\bibfnamefont {Asuka}\
  \bibnamefont {Ito}}, \ and\ \bibinfo {author} {\bibfnamefont {Jiro}\
  \bibnamefont {Soda}},\ }\bibfield  {title} {\enquote {\bibinfo {title}
  {{Pulsar timing residual induced by ultralight vector dark matter}},}\ }\href
  {\doibase 10.1140/epjc/s10052-020-7990-y} {\bibfield  {journal} {\bibinfo
  {journal} {Eur. Phys. J. C}\ }\textbf {\bibinfo {volume} {80}},\ \bibinfo
  {pages} {419} (\bibinfo {year} {2020})},\ \Eprint
  {http://arxiv.org/abs/1912.10210} {arXiv:1912.10210 [gr-qc]} \BibitemShut
  {NoStop}%
\bibitem [{\citenamefont {Aoki}\ and\ \citenamefont
  {Mukohyama}(2016)}]{Aoki:2016zgp}%
  \BibitemOpen
  \bibfield  {author} {\bibinfo {author} {\bibfnamefont {Katsuki}\ \bibnamefont
  {Aoki}}\ and\ \bibinfo {author} {\bibfnamefont {Shinji}\ \bibnamefont
  {Mukohyama}},\ }\bibfield  {title} {\enquote {\bibinfo {title} {{Massive
  gravitons as dark matter and gravitational waves}},}\ }\href {\doibase
  10.1103/PhysRevD.94.024001} {\bibfield  {journal} {\bibinfo  {journal} {Phys.
  Rev. D}\ }\textbf {\bibinfo {volume} {94}},\ \bibinfo {pages} {024001}
  (\bibinfo {year} {2016})},\ \Eprint {http://arxiv.org/abs/1604.06704}
  {arXiv:1604.06704 [hep-th]} \BibitemShut {NoStop}%
\bibitem [{\citenamefont {Babichev}\ \emph
  {et~al.}(2016{\natexlab{a}})\citenamefont {Babichev}, \citenamefont
  {Marzola}, \citenamefont {Raidal}, \citenamefont {Schmidt-May}, \citenamefont
  {Urban}, \citenamefont {Veerm\"ae},\ and\ \citenamefont {von
  Strauss}}]{Babichev:2016hir}%
  \BibitemOpen
  \bibfield  {author} {\bibinfo {author} {\bibfnamefont {Eugeny}\ \bibnamefont
  {Babichev}}, \bibinfo {author} {\bibfnamefont {Luca}\ \bibnamefont
  {Marzola}}, \bibinfo {author} {\bibfnamefont {Martti}\ \bibnamefont
  {Raidal}}, \bibinfo {author} {\bibfnamefont {Angnis}\ \bibnamefont
  {Schmidt-May}}, \bibinfo {author} {\bibfnamefont {Federico}\ \bibnamefont
  {Urban}}, \bibinfo {author} {\bibfnamefont {Hardi}\ \bibnamefont
  {Veerm\"ae}}, \ and\ \bibinfo {author} {\bibfnamefont {Mikael}\ \bibnamefont
  {von Strauss}},\ }\bibfield  {title} {\enquote {\bibinfo {title}
  {{Bigravitational origin of dark matter}},}\ }\href {\doibase
  10.1103/PhysRevD.94.084055} {\bibfield  {journal} {\bibinfo  {journal} {Phys.
  Rev. D}\ }\textbf {\bibinfo {volume} {94}},\ \bibinfo {pages} {084055}
  (\bibinfo {year} {2016}{\natexlab{a}})},\ \Eprint
  {http://arxiv.org/abs/1604.08564} {arXiv:1604.08564 [hep-ph]} \BibitemShut
  {NoStop}%
\bibitem [{\citenamefont {Babichev}\ \emph
  {et~al.}(2016{\natexlab{b}})\citenamefont {Babichev}, \citenamefont
  {Marzola}, \citenamefont {Raidal}, \citenamefont {Schmidt-May}, \citenamefont
  {Urban}, \citenamefont {Veerm\"ae},\ and\ \citenamefont {von
  Strauss}}]{Babichev:2016bxi}%
  \BibitemOpen
  \bibfield  {author} {\bibinfo {author} {\bibfnamefont {Eugeny}\ \bibnamefont
  {Babichev}}, \bibinfo {author} {\bibfnamefont {Luca}\ \bibnamefont
  {Marzola}}, \bibinfo {author} {\bibfnamefont {Martti}\ \bibnamefont
  {Raidal}}, \bibinfo {author} {\bibfnamefont {Angnis}\ \bibnamefont
  {Schmidt-May}}, \bibinfo {author} {\bibfnamefont {Federico}\ \bibnamefont
  {Urban}}, \bibinfo {author} {\bibfnamefont {Hardi}\ \bibnamefont
  {Veerm\"ae}}, \ and\ \bibinfo {author} {\bibfnamefont {Mikael}\ \bibnamefont
  {von Strauss}},\ }\bibfield  {title} {\enquote {\bibinfo {title} {{Heavy
  spin-2 Dark Matter}},}\ }\href {\doibase 10.1088/1475-7516/2016/09/016}
  {\bibfield  {journal} {\bibinfo  {journal} {JCAP}\ }\textbf {\bibinfo
  {volume} {09}},\ \bibinfo {pages} {016} (\bibinfo {year}
  {2016}{\natexlab{b}})},\ \Eprint {http://arxiv.org/abs/1607.03497}
  {arXiv:1607.03497 [hep-th]} \BibitemShut {NoStop}%
\bibitem [{\citenamefont {Aoki}\ and\ \citenamefont
  {Maeda}(2018)}]{Aoki:2017cnz}%
  \BibitemOpen
  \bibfield  {author} {\bibinfo {author} {\bibfnamefont {Katsuki}\ \bibnamefont
  {Aoki}}\ and\ \bibinfo {author} {\bibfnamefont {Kei-ichi}\ \bibnamefont
  {Maeda}},\ }\bibfield  {title} {\enquote {\bibinfo {title} {{Condensate of
  Massive Graviton and Dark Matter}},}\ }\href {\doibase
  10.1103/PhysRevD.97.044002} {\bibfield  {journal} {\bibinfo  {journal} {Phys.
  Rev. D}\ }\textbf {\bibinfo {volume} {97}},\ \bibinfo {pages} {044002}
  (\bibinfo {year} {2018})},\ \Eprint {http://arxiv.org/abs/1707.05003}
  {arXiv:1707.05003 [hep-th]} \BibitemShut {NoStop}%
\bibitem [{\citenamefont {Hassan}\ and\ \citenamefont
  {Rosen}(2012)}]{Hassan:2011zd}%
  \BibitemOpen
  \bibfield  {author} {\bibinfo {author} {\bibfnamefont {S.~F.}\ \bibnamefont
  {Hassan}}\ and\ \bibinfo {author} {\bibfnamefont {Rachel~A.}\ \bibnamefont
  {Rosen}},\ }\bibfield  {title} {\enquote {\bibinfo {title} {{Bimetric Gravity
  from Ghost-free Massive Gravity}},}\ }\href {\doibase
  10.1007/JHEP02(2012)126} {\bibfield  {journal} {\bibinfo  {journal} {JHEP}\
  }\textbf {\bibinfo {volume} {02}},\ \bibinfo {pages} {126} (\bibinfo {year}
  {2012})},\ \Eprint {http://arxiv.org/abs/1109.3515} {arXiv:1109.3515
  [hep-th]} \BibitemShut {NoStop}%
\bibitem [{\citenamefont {Marzola}\ \emph {et~al.}(2018)\citenamefont
  {Marzola}, \citenamefont {Raidal},\ and\ \citenamefont
  {Urban}}]{Marzola:2017lbt}%
  \BibitemOpen
  \bibfield  {author} {\bibinfo {author} {\bibfnamefont {Luca}\ \bibnamefont
  {Marzola}}, \bibinfo {author} {\bibfnamefont {Martti}\ \bibnamefont
  {Raidal}}, \ and\ \bibinfo {author} {\bibfnamefont {Federico~R.}\
  \bibnamefont {Urban}},\ }\bibfield  {title} {\enquote {\bibinfo {title}
  {{Oscillating Spin-2 Dark Matter}},}\ }\href {\doibase
  10.1103/PhysRevD.97.024010} {\bibfield  {journal} {\bibinfo  {journal} {Phys.
  Rev. D}\ }\textbf {\bibinfo {volume} {97}},\ \bibinfo {pages} {024010}
  (\bibinfo {year} {2018})},\ \Eprint {http://arxiv.org/abs/1708.04253}
  {arXiv:1708.04253 [hep-ph]} \BibitemShut {NoStop}%
\bibitem [{\citenamefont {Gialamas}\ and\ \citenamefont
  {Tamvakis}(2023)}]{Gialamas:2023aim}%
  \BibitemOpen
  \bibfield  {author} {\bibinfo {author} {\bibfnamefont {Ioannis~D.}\
  \bibnamefont {Gialamas}}\ and\ \bibinfo {author} {\bibfnamefont {Kyriakos}\
  \bibnamefont {Tamvakis}},\ }\bibfield  {title} {\enquote {\bibinfo {title}
  {{Bimetric-affine quadratic gravity}},}\ }\href {\doibase
  10.1103/PhysRevD.107.104012} {\bibfield  {journal} {\bibinfo  {journal}
  {Phys. Rev. D}\ }\textbf {\bibinfo {volume} {107}},\ \bibinfo {pages}
  {104012} (\bibinfo {year} {2023})},\ \Eprint
  {http://arxiv.org/abs/2303.11353} {arXiv:2303.11353 [gr-qc]} \BibitemShut
  {NoStop}%
\bibitem [{\citenamefont {Poulin}\ \emph {et~al.}(2018)\citenamefont {Poulin},
  \citenamefont {Smith}, \citenamefont {Grin}, \citenamefont {Karwal},\ and\
  \citenamefont {Kamionkowski}}]{Poulin:2018dzj}%
  \BibitemOpen
  \bibfield  {author} {\bibinfo {author} {\bibfnamefont {Vivian}\ \bibnamefont
  {Poulin}}, \bibinfo {author} {\bibfnamefont {Tristan~L.}\ \bibnamefont
  {Smith}}, \bibinfo {author} {\bibfnamefont {Daniel}\ \bibnamefont {Grin}},
  \bibinfo {author} {\bibfnamefont {Tanvi}\ \bibnamefont {Karwal}}, \ and\
  \bibinfo {author} {\bibfnamefont {Marc}\ \bibnamefont {Kamionkowski}},\
  }\bibfield  {title} {\enquote {\bibinfo {title} {{Cosmological implications
  of ultralight axionlike fields}},}\ }\href {\doibase
  10.1103/PhysRevD.98.083525} {\bibfield  {journal} {\bibinfo  {journal} {Phys.
  Rev. D}\ }\textbf {\bibinfo {volume} {98}},\ \bibinfo {pages} {083525}
  (\bibinfo {year} {2018})},\ \Eprint {http://arxiv.org/abs/1806.10608}
  {arXiv:1806.10608 [astro-ph.CO]} \BibitemShut {NoStop}%
\bibitem [{\citenamefont {Ir\v{s}i\v{c}}\ \emph {et~al.}(2017)\citenamefont
  {Ir\v{s}i\v{c}}, \citenamefont {Viel}, \citenamefont {Haehnelt},
  \citenamefont {Bolton},\ and\ \citenamefont {Becker}}]{Irsic:2017yje}%
  \BibitemOpen
  \bibfield  {author} {\bibinfo {author} {\bibfnamefont {Vid}\ \bibnamefont
  {Ir\v{s}i\v{c}}}, \bibinfo {author} {\bibfnamefont {Matteo}\ \bibnamefont
  {Viel}}, \bibinfo {author} {\bibfnamefont {Martin~G.}\ \bibnamefont
  {Haehnelt}}, \bibinfo {author} {\bibfnamefont {James~S.}\ \bibnamefont
  {Bolton}}, \ and\ \bibinfo {author} {\bibfnamefont {George~D.}\ \bibnamefont
  {Becker}},\ }\bibfield  {title} {\enquote {\bibinfo {title} {{First
  constraints on fuzzy dark matter from Lyman-$\alpha$ forest data and
  hydrodynamical simulations}},}\ }\href {\doibase
  10.1103/PhysRevLett.119.031302} {\bibfield  {journal} {\bibinfo  {journal}
  {Phys. Rev. Lett.}\ }\textbf {\bibinfo {volume} {119}},\ \bibinfo {pages}
  {031302} (\bibinfo {year} {2017})},\ \Eprint
  {http://arxiv.org/abs/1703.04683} {arXiv:1703.04683 [astro-ph.CO]}
  \BibitemShut {NoStop}%
\bibitem [{\citenamefont {\"Unal}\ \emph {et~al.}(2021)\citenamefont {\"Unal},
  \citenamefont {Pacucci},\ and\ \citenamefont {Loeb}}]{Unal:2020jiy}%
  \BibitemOpen
  \bibfield  {author} {\bibinfo {author} {\bibfnamefont {Caner}\ \bibnamefont
  {\"Unal}}, \bibinfo {author} {\bibfnamefont {Fabio}\ \bibnamefont {Pacucci}},
  \ and\ \bibinfo {author} {\bibfnamefont {Abraham}\ \bibnamefont {Loeb}},\
  }\bibfield  {title} {\enquote {\bibinfo {title} {{Properties of ultralight
  bosons from heavy quasar spins via superradiance}},}\ }\href {\doibase
  10.1088/1475-7516/2021/05/007} {\bibfield  {journal} {\bibinfo  {journal}
  {JCAP}\ }\textbf {\bibinfo {volume} {05}},\ \bibinfo {pages} {007} (\bibinfo
  {year} {2021})},\ \Eprint {http://arxiv.org/abs/2012.12790} {arXiv:2012.12790
  [hep-ph]} \BibitemShut {NoStop}%
\bibitem [{\citenamefont {Armaleo}\ \emph
  {et~al.}(2020{\natexlab{a}})\citenamefont {Armaleo}, \citenamefont
  {L\'opez~Nacir},\ and\ \citenamefont {Urban}}]{Armaleo:2019gil}%
  \BibitemOpen
  \bibfield  {author} {\bibinfo {author} {\bibfnamefont {Juan~Manuel}\
  \bibnamefont {Armaleo}}, \bibinfo {author} {\bibfnamefont {Diana}\
  \bibnamefont {L\'opez~Nacir}}, \ and\ \bibinfo {author} {\bibfnamefont
  {Federico~R.}\ \bibnamefont {Urban}},\ }\bibfield  {title} {\enquote
  {\bibinfo {title} {{Binary pulsars as probes for spin-2 ultralight dark
  matter}},}\ }\href {\doibase 10.1088/1475-7516/2020/01/053} {\bibfield
  {journal} {\bibinfo  {journal} {JCAP}\ }\textbf {\bibinfo {volume} {01}},\
  \bibinfo {pages} {053} (\bibinfo {year} {2020}{\natexlab{a}})},\ \Eprint
  {http://arxiv.org/abs/1909.13814} {arXiv:1909.13814 [astro-ph.HE]}
  \BibitemShut {NoStop}%
\bibitem [{\citenamefont {Blas}\ \emph {et~al.}(2020)\citenamefont {Blas},
  \citenamefont {L\'opez~Nacir},\ and\ \citenamefont
  {Sibiryakov}}]{Blas:2019hxz}%
  \BibitemOpen
  \bibfield  {author} {\bibinfo {author} {\bibfnamefont {Diego}\ \bibnamefont
  {Blas}}, \bibinfo {author} {\bibfnamefont {Diana}\ \bibnamefont
  {L\'opez~Nacir}}, \ and\ \bibinfo {author} {\bibfnamefont {Sergey}\
  \bibnamefont {Sibiryakov}},\ }\bibfield  {title} {\enquote {\bibinfo {title}
  {{Secular effects of ultralight dark matter on binary pulsars}},}\ }\href
  {\doibase 10.1103/PhysRevD.101.063016} {\bibfield  {journal} {\bibinfo
  {journal} {Phys. Rev. D}\ }\textbf {\bibinfo {volume} {101}},\ \bibinfo
  {pages} {063016} (\bibinfo {year} {2020})},\ \Eprint
  {http://arxiv.org/abs/1910.08544} {arXiv:1910.08544 [gr-qc]} \BibitemShut
  {NoStop}%
\bibitem [{\citenamefont {Porayko}\ and\ \citenamefont
  {Postnov}(2014)}]{Porayko:2014rfa}%
  \BibitemOpen
  \bibfield  {author} {\bibinfo {author} {\bibfnamefont {N.~K.}\ \bibnamefont
  {Porayko}}\ and\ \bibinfo {author} {\bibfnamefont {K.~A.}\ \bibnamefont
  {Postnov}},\ }\bibfield  {title} {\enquote {\bibinfo {title} {{Constraints on
  ultralight scalar dark matter from pulsar timing}},}\ }\href {\doibase
  10.1103/PhysRevD.90.062008} {\bibfield  {journal} {\bibinfo  {journal} {Phys.
  Rev. D}\ }\textbf {\bibinfo {volume} {90}},\ \bibinfo {pages} {062008}
  (\bibinfo {year} {2014})},\ \Eprint {http://arxiv.org/abs/1408.4670}
  {arXiv:1408.4670 [astro-ph.CO]} \BibitemShut {NoStop}%
\bibitem [{\citenamefont {Porayko}\ \emph {et~al.}(2018)\citenamefont {Porayko}
  \emph {et~al.}}]{Porayko:2018sfa}%
  \BibitemOpen
  \bibfield  {author} {\bibinfo {author} {\bibfnamefont {Nataliya~K.}\
  \bibnamefont {Porayko}} \emph {et~al.},\ }\bibfield  {title} {\enquote
  {\bibinfo {title} {{Parkes Pulsar Timing Array constraints on ultralight
  scalar-field dark matter}},}\ }\href {\doibase 10.1103/PhysRevD.98.102002}
  {\bibfield  {journal} {\bibinfo  {journal} {Phys. Rev. D}\ }\textbf {\bibinfo
  {volume} {98}},\ \bibinfo {pages} {102002} (\bibinfo {year} {2018})},\
  \Eprint {http://arxiv.org/abs/1810.03227} {arXiv:1810.03227 [astro-ph.CO]}
  \BibitemShut {NoStop}%
\bibitem [{\citenamefont {Kato}\ and\ \citenamefont
  {Soda}(2020)}]{Kato:2019bqz}%
  \BibitemOpen
  \bibfield  {author} {\bibinfo {author} {\bibfnamefont {Ryo}\ \bibnamefont
  {Kato}}\ and\ \bibinfo {author} {\bibfnamefont {Jiro}\ \bibnamefont {Soda}},\
  }\bibfield  {title} {\enquote {\bibinfo {title} {{Search for ultralight
  scalar dark matter with NANOGrav pulsar timing arrays}},}\ }\href {\doibase
  10.1088/1475-7516/2020/09/036} {\bibfield  {journal} {\bibinfo  {journal}
  {JCAP}\ }\textbf {\bibinfo {volume} {09}},\ \bibinfo {pages} {036} (\bibinfo
  {year} {2020})},\ \Eprint {http://arxiv.org/abs/1904.09143} {arXiv:1904.09143
  [astro-ph.HE]} \BibitemShut {NoStop}%
\bibitem [{\citenamefont {Xue}\ \emph {et~al.}(2022)\citenamefont {Xue} \emph
  {et~al.}}]{PPTA:2021uzb}%
  \BibitemOpen
  \bibfield  {author} {\bibinfo {author} {\bibfnamefont {Xiao}\ \bibnamefont
  {Xue}} \emph {et~al.} (\bibinfo {collaboration} {PPTA}),\ }\bibfield  {title}
  {\enquote {\bibinfo {title} {{High-precision search for dark photon dark
  matter with the Parkes Pulsar Timing Array}},}\ }\href {\doibase
  10.1103/PhysRevResearch.4.L012022} {\bibfield  {journal} {\bibinfo  {journal}
  {Phys. Rev. Res.}\ }\textbf {\bibinfo {volume} {4}},\ \bibinfo {pages}
  {L012022} (\bibinfo {year} {2022})},\ \Eprint
  {http://arxiv.org/abs/2112.07687} {arXiv:2112.07687 [hep-ph]} \BibitemShut
  {NoStop}%
\bibitem [{\citenamefont {Wu}\ \emph {et~al.}(2022)\citenamefont {Wu},
  \citenamefont {Chen}, \citenamefont {Huang}, \citenamefont {Zhu},
  \citenamefont {Bhat}, \citenamefont {Feng}, \citenamefont {Hobbs},
  \citenamefont {Manchester}, \citenamefont {Russell},\ and\ \citenamefont
  {Shannon}}]{PPTA:2022eul}%
  \BibitemOpen
  \bibfield  {author} {\bibinfo {author} {\bibfnamefont {Yu-Mei}\ \bibnamefont
  {Wu}}, \bibinfo {author} {\bibfnamefont {Zu-Cheng}\ \bibnamefont {Chen}},
  \bibinfo {author} {\bibfnamefont {Qing-Guo}\ \bibnamefont {Huang}}, \bibinfo
  {author} {\bibfnamefont {Xingjiang}\ \bibnamefont {Zhu}}, \bibinfo {author}
  {\bibfnamefont {N.~D.~Ramesh}\ \bibnamefont {Bhat}}, \bibinfo {author}
  {\bibfnamefont {Yi}~\bibnamefont {Feng}}, \bibinfo {author} {\bibfnamefont
  {George}\ \bibnamefont {Hobbs}}, \bibinfo {author} {\bibfnamefont
  {Richard~N.}\ \bibnamefont {Manchester}}, \bibinfo {author} {\bibfnamefont
  {Christopher~J.}\ \bibnamefont {Russell}}, \ and\ \bibinfo {author}
  {\bibfnamefont {R.~M.}\ \bibnamefont {Shannon}} (\bibinfo {collaboration}
  {PPTA}),\ }\bibfield  {title} {\enquote {\bibinfo {title} {{Constraining
  ultralight vector dark matter with the Parkes Pulsar Timing Array second data
  release}},}\ }\href {\doibase 10.1103/PhysRevD.106.L081101} {\bibfield
  {journal} {\bibinfo  {journal} {Phys. Rev. D}\ }\textbf {\bibinfo {volume}
  {106}},\ \bibinfo {pages} {L081101} (\bibinfo {year} {2022})},\ \Eprint
  {http://arxiv.org/abs/2210.03880} {arXiv:2210.03880 [astro-ph.CO]}
  \BibitemShut {NoStop}%
\bibitem [{\citenamefont {Armaleo}\ \emph
  {et~al.}(2020{\natexlab{b}})\citenamefont {Armaleo}, \citenamefont
  {L\'opez~Nacir},\ and\ \citenamefont {Urban}}]{Armaleo:2020yml}%
  \BibitemOpen
  \bibfield  {author} {\bibinfo {author} {\bibfnamefont {Juan~Manuel}\
  \bibnamefont {Armaleo}}, \bibinfo {author} {\bibfnamefont {Diana}\
  \bibnamefont {L\'opez~Nacir}}, \ and\ \bibinfo {author} {\bibfnamefont
  {Federico~R.}\ \bibnamefont {Urban}},\ }\bibfield  {title} {\enquote
  {\bibinfo {title} {{Pulsar timing array constraints on spin-2 ULDM}},}\
  }\href {\doibase 10.1088/1475-7516/2020/09/031} {\bibfield  {journal}
  {\bibinfo  {journal} {JCAP}\ }\textbf {\bibinfo {volume} {09}},\ \bibinfo
  {pages} {031} (\bibinfo {year} {2020}{\natexlab{b}})},\ \Eprint
  {http://arxiv.org/abs/2005.03731} {arXiv:2005.03731 [astro-ph.CO]}
  \BibitemShut {NoStop}%
\bibitem [{\citenamefont {{Sazhin}}(1978)}]{1978SvA....22...36S}%
  \BibitemOpen
  \bibfield  {author} {\bibinfo {author} {\bibfnamefont {M.~V.}\ \bibnamefont
  {{Sazhin}}},\ }\bibfield  {title} {\enquote {\bibinfo {title} {{Opportunities
  for detecting ultralong gravitational waves}},}\ }\href@noop {} {\bibfield
  {journal} {\bibinfo  {journal} {\sovast}\ }\textbf {\bibinfo {volume} {22}},\
  \bibinfo {pages} {36--38} (\bibinfo {year} {1978})}\BibitemShut {NoStop}%
\bibitem [{\citenamefont {Detweiler}(1979)}]{Detweiler:1979wn}%
  \BibitemOpen
  \bibfield  {author} {\bibinfo {author} {\bibfnamefont {Steven~L.}\
  \bibnamefont {Detweiler}},\ }\bibfield  {title} {\enquote {\bibinfo {title}
  {{Pulsar timing measurements and the search for gravitational waves}},}\
  }\href {\doibase 10.1086/157593} {\bibfield  {journal} {\bibinfo  {journal}
  {Astrophys. J.}\ }\textbf {\bibinfo {volume} {234}},\ \bibinfo {pages}
  {1100--1104} (\bibinfo {year} {1979})}\BibitemShut {NoStop}%
\bibitem [{\citenamefont {{Foster}}\ and\ \citenamefont
  {{Backer}}(1990)}]{1990ApJ...361..300F}%
  \BibitemOpen
  \bibfield  {author} {\bibinfo {author} {\bibfnamefont {R.~S.}\ \bibnamefont
  {{Foster}}}\ and\ \bibinfo {author} {\bibfnamefont {D.~C.}\ \bibnamefont
  {{Backer}}},\ }\bibfield  {title} {\enquote {\bibinfo {title} {{Constructing
  a Pulsar Timing Array}},}\ }\href {\doibase 10.1086/169195} {\bibfield
  {journal} {\bibinfo  {journal} {\apj}\ }\textbf {\bibinfo {volume} {361}},\
  \bibinfo {pages} {300} (\bibinfo {year} {1990})}\BibitemShut {NoStop}%
\bibitem [{\citenamefont {Kaplan}\ \emph {et~al.}(2022)\citenamefont {Kaplan},
  \citenamefont {Mitridate},\ and\ \citenamefont {Trickle}}]{Kaplan:2022lmz}%
  \BibitemOpen
  \bibfield  {author} {\bibinfo {author} {\bibfnamefont {David~E.}\
  \bibnamefont {Kaplan}}, \bibinfo {author} {\bibfnamefont {Andrea}\
  \bibnamefont {Mitridate}}, \ and\ \bibinfo {author} {\bibfnamefont {Tanner}\
  \bibnamefont {Trickle}},\ }\bibfield  {title} {\enquote {\bibinfo {title}
  {{Constraining fundamental constant variations from ultralight dark matter
  with pulsar timing arrays}},}\ }\href {\doibase 10.1103/PhysRevD.106.035032}
  {\bibfield  {journal} {\bibinfo  {journal} {Phys. Rev. D}\ }\textbf {\bibinfo
  {volume} {106}},\ \bibinfo {pages} {035032} (\bibinfo {year} {2022})},\
  \Eprint {http://arxiv.org/abs/2205.06817} {arXiv:2205.06817 [hep-ph]}
  \BibitemShut {NoStop}%
\bibitem [{\citenamefont {Sun}\ \emph {et~al.}(2022)\citenamefont {Sun},
  \citenamefont {Yang},\ and\ \citenamefont {Zhang}}]{Sun:2021yra}%
  \BibitemOpen
  \bibfield  {author} {\bibinfo {author} {\bibfnamefont {Sichun}\ \bibnamefont
  {Sun}}, \bibinfo {author} {\bibfnamefont {Xing-Yu}\ \bibnamefont {Yang}}, \
  and\ \bibinfo {author} {\bibfnamefont {Yun-Long}\ \bibnamefont {Zhang}},\
  }\bibfield  {title} {\enquote {\bibinfo {title} {{Pulsar timing residual
  induced by wideband ultralight dark matter with spin 0,1,2}},}\ }\href
  {\doibase 10.1103/PhysRevD.106.066006} {\bibfield  {journal} {\bibinfo
  {journal} {Phys. Rev. D}\ }\textbf {\bibinfo {volume} {106}},\ \bibinfo
  {pages} {066006} (\bibinfo {year} {2022})},\ \Eprint
  {http://arxiv.org/abs/2112.15593} {arXiv:2112.15593 [astro-ph.CO]}
  \BibitemShut {NoStop}%
\bibitem [{\citenamefont {Unal}\ \emph {et~al.}(2022)\citenamefont {Unal},
  \citenamefont {Urban},\ and\ \citenamefont {Kovetz}}]{Unal:2022ooa}%
  \BibitemOpen
  \bibfield  {author} {\bibinfo {author} {\bibfnamefont {Caner}\ \bibnamefont
  {Unal}}, \bibinfo {author} {\bibfnamefont {Federico~R.}\ \bibnamefont
  {Urban}}, \ and\ \bibinfo {author} {\bibfnamefont {Ely~D.}\ \bibnamefont
  {Kovetz}},\ }\bibfield  {title} {\enquote {\bibinfo {title} {{Probing
  ultralight scalar, vector and tensor dark matter with pulsar timing
  arrays}},}\ }\href@noop {} {\  (\bibinfo {year} {2022})},\ \Eprint
  {http://arxiv.org/abs/2209.02741} {arXiv:2209.02741 [astro-ph.CO]}
  \BibitemShut {NoStop}%
\bibitem [{\citenamefont {Kolb}\ \emph {et~al.}(2023)\citenamefont {Kolb},
  \citenamefont {Ling}, \citenamefont {Long},\ and\ \citenamefont
  {Rosen}}]{Kolb:2023dzp}%
  \BibitemOpen
  \bibfield  {author} {\bibinfo {author} {\bibfnamefont {Edward~W.}\
  \bibnamefont {Kolb}}, \bibinfo {author} {\bibfnamefont {Siyang}\ \bibnamefont
  {Ling}}, \bibinfo {author} {\bibfnamefont {Andrew~J.}\ \bibnamefont {Long}},
  \ and\ \bibinfo {author} {\bibfnamefont {Rachel~A.}\ \bibnamefont {Rosen}},\
  }\bibfield  {title} {\enquote {\bibinfo {title} {{Cosmological gravitational
  particle production of massive spin-2 particles}},}\ }\href@noop {} {\
  (\bibinfo {year} {2023})},\ \Eprint {http://arxiv.org/abs/2302.04390}
  {arXiv:2302.04390 [astro-ph.CO]} \BibitemShut {NoStop}%
\bibitem [{\citenamefont {Graham}\ \emph {et~al.}(2016)\citenamefont {Graham},
  \citenamefont {Mardon},\ and\ \citenamefont {Rajendran}}]{Graham:2015rva}%
  \BibitemOpen
  \bibfield  {author} {\bibinfo {author} {\bibfnamefont {Peter~W.}\
  \bibnamefont {Graham}}, \bibinfo {author} {\bibfnamefont {Jeremy}\
  \bibnamefont {Mardon}}, \ and\ \bibinfo {author} {\bibfnamefont {Surjeet}\
  \bibnamefont {Rajendran}},\ }\bibfield  {title} {\enquote {\bibinfo {title}
  {{Vector Dark Matter from Inflationary Fluctuations}},}\ }\href {\doibase
  10.1103/PhysRevD.93.103520} {\bibfield  {journal} {\bibinfo  {journal} {Phys.
  Rev. D}\ }\textbf {\bibinfo {volume} {93}},\ \bibinfo {pages} {103520}
  (\bibinfo {year} {2016})},\ \Eprint {http://arxiv.org/abs/1504.02102}
  {arXiv:1504.02102 [hep-ph]} \BibitemShut {NoStop}%
\bibitem [{\citenamefont {Salucci}\ \emph {et~al.}(2010)\citenamefont
  {Salucci}, \citenamefont {Nesti}, \citenamefont {Gentile},\ and\
  \citenamefont {Martins}}]{Salucci:2010qr}%
  \BibitemOpen
  \bibfield  {author} {\bibinfo {author} {\bibfnamefont {P.}~\bibnamefont
  {Salucci}}, \bibinfo {author} {\bibfnamefont {F.}~\bibnamefont {Nesti}},
  \bibinfo {author} {\bibfnamefont {G.}~\bibnamefont {Gentile}}, \ and\
  \bibinfo {author} {\bibfnamefont {C.~F.}\ \bibnamefont {Martins}},\
  }\bibfield  {title} {\enquote {\bibinfo {title} {{The dark matter density at
  the Sun's location}},}\ }\href {\doibase 10.1051/0004-6361/201014385}
  {\bibfield  {journal} {\bibinfo  {journal} {Astron. Astrophys.}\ }\textbf
  {\bibinfo {volume} {523}},\ \bibinfo {pages} {A83} (\bibinfo {year}
  {2010})},\ \Eprint {http://arxiv.org/abs/1003.3101} {arXiv:1003.3101
  [astro-ph.GA]} \BibitemShut {NoStop}%
\bibitem [{\citenamefont {Hlozek}\ \emph {et~al.}(2018)\citenamefont {Hlozek},
  \citenamefont {Marsh},\ and\ \citenamefont {Grin}}]{Hlozek:2017zzf}%
  \BibitemOpen
  \bibfield  {author} {\bibinfo {author} {\bibfnamefont {Ren\'ee}\ \bibnamefont
  {Hlozek}}, \bibinfo {author} {\bibfnamefont {David J.~E.}\ \bibnamefont
  {Marsh}}, \ and\ \bibinfo {author} {\bibfnamefont {Daniel}\ \bibnamefont
  {Grin}},\ }\bibfield  {title} {\enquote {\bibinfo {title} {{Using the Full
  Power of the Cosmic Microwave Background to Probe Axion Dark Matter}},}\
  }\href {\doibase 10.1093/mnras/sty271} {\bibfield  {journal} {\bibinfo
  {journal} {Mon. Not. Roy. Astron. Soc.}\ }\textbf {\bibinfo {volume} {476}},\
  \bibinfo {pages} {3063--3085} (\bibinfo {year} {2018})},\ \Eprint
  {http://arxiv.org/abs/1708.05681} {arXiv:1708.05681 [astro-ph.CO]}
  \BibitemShut {NoStop}%
\bibitem [{\citenamefont {Bar}\ \emph {et~al.}(2018)\citenamefont {Bar},
  \citenamefont {Blas}, \citenamefont {Blum},\ and\ \citenamefont
  {Sibiryakov}}]{Bar:2018acw}%
  \BibitemOpen
  \bibfield  {author} {\bibinfo {author} {\bibfnamefont {Nitsan}\ \bibnamefont
  {Bar}}, \bibinfo {author} {\bibfnamefont {Diego}\ \bibnamefont {Blas}},
  \bibinfo {author} {\bibfnamefont {Kfir}\ \bibnamefont {Blum}}, \ and\
  \bibinfo {author} {\bibfnamefont {Sergey}\ \bibnamefont {Sibiryakov}},\
  }\bibfield  {title} {\enquote {\bibinfo {title} {{Galactic rotation curves
  versus ultralight dark matter: Implications of the soliton-host halo
  relation}},}\ }\href {\doibase 10.1103/PhysRevD.98.083027} {\bibfield
  {journal} {\bibinfo  {journal} {Phys. Rev. D}\ }\textbf {\bibinfo {volume}
  {98}},\ \bibinfo {pages} {083027} (\bibinfo {year} {2018})},\ \Eprint
  {http://arxiv.org/abs/1805.00122} {arXiv:1805.00122 [astro-ph.CO]}
  \BibitemShut {NoStop}%
\bibitem [{\citenamefont {Nadler}\ \emph {et~al.}(2019)\citenamefont {Nadler},
  \citenamefont {Gluscevic}, \citenamefont {Boddy},\ and\ \citenamefont
  {Wechsler}}]{Nadler:2019zrb}%
  \BibitemOpen
  \bibfield  {author} {\bibinfo {author} {\bibfnamefont {Ethan~O.}\
  \bibnamefont {Nadler}}, \bibinfo {author} {\bibfnamefont {Vera}\ \bibnamefont
  {Gluscevic}}, \bibinfo {author} {\bibfnamefont {Kimberly~K.}\ \bibnamefont
  {Boddy}}, \ and\ \bibinfo {author} {\bibfnamefont {Risa~H.}\ \bibnamefont
  {Wechsler}},\ }\bibfield  {title} {\enquote {\bibinfo {title} {{Constraints
  on Dark Matter Microphysics from the Milky Way Satellite Population}},}\
  }\href {\doibase 10.3847/2041-8213/ab1eb2} {\bibfield  {journal} {\bibinfo
  {journal} {Astrophys. J. Lett.}\ }\textbf {\bibinfo {volume} {878}},\
  \bibinfo {pages} {32} (\bibinfo {year} {2019})},\ \bibinfo {note} {[Erratum:
  Astrophys.J.Lett. 897, L46 (2020), Erratum: Astrophys.J. 897, L46 (2020)]},\
  \Eprint {http://arxiv.org/abs/1904.10000} {arXiv:1904.10000 [astro-ph.CO]}
  \BibitemShut {NoStop}%
\bibitem [{\citenamefont {Rogers}\ and\ \citenamefont
  {Peiris}(2021)}]{Rogers:2020ltq}%
  \BibitemOpen
  \bibfield  {author} {\bibinfo {author} {\bibfnamefont {Keir~K.}\ \bibnamefont
  {Rogers}}\ and\ \bibinfo {author} {\bibfnamefont {Hiranya~V.}\ \bibnamefont
  {Peiris}},\ }\bibfield  {title} {\enquote {\bibinfo {title} {{Strong Bound on
  Canonical Ultralight Axion Dark Matter from the Lyman-Alpha Forest}},}\
  }\href {\doibase 10.1103/PhysRevLett.126.071302} {\bibfield  {journal}
  {\bibinfo  {journal} {Phys. Rev. Lett.}\ }\textbf {\bibinfo {volume} {126}},\
  \bibinfo {pages} {071302} (\bibinfo {year} {2021})},\ \Eprint
  {http://arxiv.org/abs/2007.12705} {arXiv:2007.12705 [astro-ph.CO]}
  \BibitemShut {NoStop}%
\bibitem [{\citenamefont {Gonz\'alez-Morales}\ \emph
  {et~al.}(2017)\citenamefont {Gonz\'alez-Morales}, \citenamefont {Marsh},
  \citenamefont {Pe\~narrubia},\ and\ \citenamefont {Ure\~na
  L\'opez}}]{Gonzalez-Morales:2016yaf}%
  \BibitemOpen
  \bibfield  {author} {\bibinfo {author} {\bibfnamefont {Alma~X.}\ \bibnamefont
  {Gonz\'alez-Morales}}, \bibinfo {author} {\bibfnamefont {David J.~E.}\
  \bibnamefont {Marsh}}, \bibinfo {author} {\bibfnamefont {Jorge}\ \bibnamefont
  {Pe\~narrubia}}, \ and\ \bibinfo {author} {\bibfnamefont {Luis~A.}\
  \bibnamefont {Ure\~na L\'opez}},\ }\bibfield  {title} {\enquote {\bibinfo
  {title} {{Unbiased constraints on ultralight axion mass from dwarf spheroidal
  galaxies}},}\ }\href {\doibase 10.1093/mnras/stx1941} {\bibfield  {journal}
  {\bibinfo  {journal} {Mon. Not. Roy. Astron. Soc.}\ }\textbf {\bibinfo
  {volume} {472}},\ \bibinfo {pages} {1346--1360} (\bibinfo {year} {2017})},\
  \Eprint {http://arxiv.org/abs/1609.05856} {arXiv:1609.05856 [astro-ph.CO]}
  \BibitemShut {NoStop}%
\bibitem [{\citenamefont {Kendall}\ and\ \citenamefont
  {Easther}(2020)}]{Kendall:2019fep}%
  \BibitemOpen
  \bibfield  {author} {\bibinfo {author} {\bibfnamefont {Emily}\ \bibnamefont
  {Kendall}}\ and\ \bibinfo {author} {\bibfnamefont {Richard}\ \bibnamefont
  {Easther}},\ }\bibfield  {title} {\enquote {\bibinfo {title} {{The Core-Cusp
  Problem Revisited: ULDM vs. CDM}},}\ }\href {\doibase 10.1017/pasa.2020.3}
  {\bibfield  {journal} {\bibinfo  {journal} {Publ. Astron. Soc. Austral.}\
  }\textbf {\bibinfo {volume} {37}},\ \bibinfo {pages} {e009} (\bibinfo {year}
  {2020})},\ \Eprint {http://arxiv.org/abs/1908.02508} {arXiv:1908.02508
  [astro-ph.CO]} \BibitemShut {NoStop}%
\bibitem [{\citenamefont {Zhu}\ \emph {et~al.}(2014)\citenamefont {Zhu} \emph
  {et~al.}}]{Zhu:2014rta}%
  \BibitemOpen
  \bibfield  {author} {\bibinfo {author} {\bibfnamefont {X.~J.}\ \bibnamefont
  {Zhu}} \emph {et~al.},\ }\bibfield  {title} {\enquote {\bibinfo {title} {{An
  all-sky search for continuous gravitational waves in the Parkes Pulsar Timing
  Array data set}},}\ }\href {\doibase 10.1093/mnras/stu1717} {\bibfield
  {journal} {\bibinfo  {journal} {Mon. Not. Roy. Astron. Soc.}\ }\textbf
  {\bibinfo {volume} {444}},\ \bibinfo {pages} {3709--3720} (\bibinfo {year}
  {2014})},\ \Eprint {http://arxiv.org/abs/1408.5129} {arXiv:1408.5129
  [astro-ph.GA]} \BibitemShut {NoStop}%
\bibitem [{\citenamefont {Falxa}\ \emph {et~al.}(2023)\citenamefont {Falxa}
  \emph {et~al.}}]{IPTA:2023ero}%
  \BibitemOpen
  \bibfield  {author} {\bibinfo {author} {\bibfnamefont {M.}~\bibnamefont
  {Falxa}} \emph {et~al.} (\bibinfo {collaboration} {IPTA}),\ }\bibfield
  {title} {\enquote {\bibinfo {title} {{Searching for continuous Gravitational
  Waves in the second data release of the International Pulsar Timing
  Array}},}\ }\href {\doibase 10.1093/mnras/stad812} {\bibfield  {journal}
  {\bibinfo  {journal} {Mon. Not. Roy. Astron. Soc.}\ }\textbf {\bibinfo
  {volume} {521}},\ \bibinfo {pages} {5077--5086} (\bibinfo {year} {2023})},\
  \Eprint {http://arxiv.org/abs/2303.10767} {arXiv:2303.10767 [gr-qc]}
  \BibitemShut {NoStop}%
\bibitem [{\citenamefont {Kaiser}\ \emph {et~al.}(2022)\citenamefont {Kaiser},
  \citenamefont {Pol}, \citenamefont {McLaughlin}, \citenamefont {Chen},
  \citenamefont {Hazboun}, \citenamefont {Kelley}, \citenamefont {Simon},
  \citenamefont {Taylor}, \citenamefont {Vigeland},\ and\ \citenamefont
  {Witt}}]{Kaiser:2022cma}%
  \BibitemOpen
  \bibfield  {author} {\bibinfo {author} {\bibfnamefont {Andrew~R.}\
  \bibnamefont {Kaiser}}, \bibinfo {author} {\bibfnamefont {Nihan~S.}\
  \bibnamefont {Pol}}, \bibinfo {author} {\bibfnamefont {Maura~A.}\
  \bibnamefont {McLaughlin}}, \bibinfo {author} {\bibfnamefont {Siyuan}\
  \bibnamefont {Chen}}, \bibinfo {author} {\bibfnamefont {Jeffrey~S.}\
  \bibnamefont {Hazboun}}, \bibinfo {author} {\bibfnamefont {Luke~Zoltan}\
  \bibnamefont {Kelley}}, \bibinfo {author} {\bibfnamefont {Joseph}\
  \bibnamefont {Simon}}, \bibinfo {author} {\bibfnamefont {Stephen~R.}\
  \bibnamefont {Taylor}}, \bibinfo {author} {\bibfnamefont {Sarah~J.}\
  \bibnamefont {Vigeland}}, \ and\ \bibinfo {author} {\bibfnamefont
  {Caitlin~A.}\ \bibnamefont {Witt}},\ }\bibfield  {title} {\enquote {\bibinfo
  {title} {{Disentangling Multiple Stochastic Gravitational Wave Background
  Sources in PTA Data Sets}},}\ }\href {\doibase 10.3847/1538-4357/ac86cc}
  {\bibfield  {journal} {\bibinfo  {journal} {Astrophys. J.}\ }\textbf
  {\bibinfo {volume} {938}},\ \bibinfo {pages} {115} (\bibinfo {year}
  {2022})},\ \Eprint {http://arxiv.org/abs/2208.02307} {arXiv:2208.02307
  [astro-ph.CO]} \BibitemShut {NoStop}%
\bibitem [{\citenamefont {Hellings}\ and\ \citenamefont
  {Downs}(1983)}]{Hellings:1983fr}%
  \BibitemOpen
  \bibfield  {author} {\bibinfo {author} {\bibfnamefont {R.~w.}\ \bibnamefont
  {Hellings}}\ and\ \bibinfo {author} {\bibfnamefont {G.~s.}\ \bibnamefont
  {Downs}},\ }\bibfield  {title} {\enquote {\bibinfo {title} {{UPPER LIMITS ON
  THE ISOTROPIC GRAVITATIONAL RADIATION BACKGROUND FROM PULSAR TIMING
  ANALYSIS}},}\ }\href {\doibase 10.1086/183954} {\bibfield  {journal}
  {\bibinfo  {journal} {Astrophys. J.}\ }\textbf {\bibinfo {volume} {265}},\
  \bibinfo {pages} {L39--L42} (\bibinfo {year} {1983})}\BibitemShut {NoStop}%
\bibitem [{\citenamefont {Liang}\ and\ \citenamefont
  {Trodden}(2021)}]{Liang:2021bct}%
  \BibitemOpen
  \bibfield  {author} {\bibinfo {author} {\bibfnamefont {Qiuyue}\ \bibnamefont
  {Liang}}\ and\ \bibinfo {author} {\bibfnamefont {Mark}\ \bibnamefont
  {Trodden}},\ }\bibfield  {title} {\enquote {\bibinfo {title} {{Detecting the
  stochastic gravitational wave background from massive gravity with pulsar
  timing arrays}},}\ }\href {\doibase 10.1103/PhysRevD.104.084052} {\bibfield
  {journal} {\bibinfo  {journal} {Phys. Rev. D}\ }\textbf {\bibinfo {volume}
  {104}},\ \bibinfo {pages} {084052} (\bibinfo {year} {2021})},\ \Eprint
  {http://arxiv.org/abs/2108.05344} {arXiv:2108.05344 [astro-ph.CO]}
  \BibitemShut {NoStop}%
\end{thebibliography}%

\end{document}